\documentclass[aps,prl,twocolumn,floatfix,reprint]{revtex4-1}
\usepackage{graphicx}
\usepackage{color}
\usepackage{amssymb} 
\usepackage{amsmath} 
\usepackage{natbib}
\usepackage{soul}
\usepackage{titlesec}
\usepackage{physics}
\usepackage[utf8]{inputenc}
\usepackage{hyperref}
\hypersetup{
    colorlinks=true,
    linkcolor=blue,
    filecolor=magenta,      
    urlcolor=cyan,
    citecolor=blue,
}

\setcitestyle{super, numbers, square,}

\usepackage{color}
\definecolor{pu}{rgb}{0,0.,0.55}  
\definecolor{rd}{rgb}{1,0,0} 
\definecolor{gr}{rgb}{0.6,0.6,0.6}

\begin{document}
\title{Novel characterisation of dopant-based qubits}

\author{B. Voisin$^{1,2}$}
\author{J. Salfi$^3$}
\author{R. Rahman$^{2}$}
\author{S. Rogge$^4$}

\affiliation{{$^1$ Silicon Quantum Computing,{\space}Sydney,{\space}NSW 2052,{\space}Australia}\\
{$^2$ School of Physics, The University of New South Wales, Sydney, NSW 2052, Australia}\\
{$^3$ Stewart Blusson Quantum Matter Institute, Department of Electrical and Computer Engineering, University of British Columbia, Vancouver, BC V6T 1Z4, Canada}\\
{$^4$\, Centre for Quantum Computation and Communication Technology, School of Physics, The University of New South Wales, Sydney, 2052, NSW, Australia} \\}

\begin{abstract}
Silicon is a leading qubit platform thanks to the exceptional coherence times that can be achieved and to the available commercial manufacturing platform for integration. Building scalable quantum processing architectures relies on accurate quantum state manipulation, which can only be achieved through a complete understanding of the underlying quantum state properties. This article reviews the electrical methods that have been developed to probe the quantum states encoded in individual and interacting atom qubits in silicon, from the pioneering single electron tunneling spectroscopy framework in nanoscale transistors, to radio frequency reflectometry to probe coherence properties and scanning tunneling microscopy to directly image the wave function at the atomic scale. Together with the development of atomistic simulations of realistic devices, these methods are today applied to other emerging dopant and optically addressable defect states to accelerate the engineering of quantum technologies in silicon.
\end{abstract}

\maketitle

\section{Introduction}
\label{sec:intro}

The concept of using a quantum mechanical system to overcome classical computation and simulation capabilities dates to Feynman in the early 1980s \cite{Fey82a}. Among physical systems developed to ultimately achieve quantum processing at large scale\cite{Cir95a,Bla04a,Los98a,Kan98a}, silicon has become a dominant semiconductor-based platform, fulfilling the elementary Di Vincenzo criteria \cite{DiV00a} and benefiting from long coherence times in isotopically enriched material. Three main routes have been pursued to create spin qubits in silicon. There are devices based on quantum dots in SiGe quantum wells \cite{Wat18a} or metal-oxide semiconductor devices\cite{Vel15a} and phosphorus donors \cite{Muh14a,He19a}. Kane's proposal~\cite{Kan98a} to use the nuclear spin of a phosphorus atom in silicon sparked formidable scientific and technical efforts to address and manipulate individual dopants. At that time, dopants in semiconductors were already identified as unintentional transport channels in heterostructure devices \cite{Del92a} and as the source of variability in industrial nanoscale transistor devices \cite{Ase03a}. Building on this, multiple research groups demonstrated controlled electrical transport through single dopants \cite{Sel06a,Kha07a,Pie09a,Tan10a} within 10 years after the initial proposal. This addressability was a first step to understand how dopant atoms interact with a semiconducting device environment to explore the possibility of scalable quantum technologies based on atomic systems. In parallel to this effort, donor devices have been specifically designed for quantum applications, either by ion implantation \cite{Muh14a} or by atomic precision lithography \cite{He19a}.\\

It is essential for any semiconductor-based platform to be supported by characterization tools of the host material, the qubits and their interactions. These tools, alongside the development of scaled-up devices, ensure that high qubit performance is maintained toward fault-tolerant quantum processing. In the case of dopants in silicon, some of the key challenges include maintaining a low level of spin and charge noise, which is mainly a material control challenge, in addition to mastering exchange coupling between the donors. The latter is linked to the length scale of the donor wave function with a Bohr orbit on the length scale of 1\,nm and the nonmonotonic dependence of the coupling strength between two donors with distance due to the indirect band gap of silicon. Designing robust two qubit gates is therefore nontrivial, since the wave function overlap of two donors defines the interaction strength.\\ 

Here, we review the characterization techniques that have been developed across the last decades to tackle these challenges. Importantly, these techniques go hand in hand with the engineering of devices with optimized capabilities. The spectroscopy of donors obtained from single-electron transport in nanoscale transistors can be complemented with atomistic device models \cite{Lan08a}, to link the spectral properties of single donors to their precise position in a complex device geometry. In 2014, this comparison between experiment and theory was extended to also include the charge density of the dopants, which is extracted by scanning tunnelling spectroscopy \cite{Sal14a}. For the first time, a model validation was not only based on spectroscopy but directly on the wave function level (i.e., an atom-resolved verification of the effective Schrodinger equation of a donor atom embedded in the silicon lattice). This joint experimental and wave function modelling was extended to multidopant systems, in the context of both device transport and quantum state imaging, to notably lead to detailed insights into two-dopant interactions. Finally, radio frequency techniques, such as reflectometry that eliminates the need of an external charge sensor, were developed to improve dopant addressability and readout as well as to probe coherence properties. This article closes on the possibility to extend these characterizing techniques to other sought after qubit systems in silicon, such as optically addressable rare-earth atoms or acceptor atoms with a sizeable spin-orbit coupling.

\section{Electrical detection of atomic states in nanoscale transistors}
\label{sec:trans}

The first signatures of direct transport though single atoms in nanoscale transistors~\cite{Sel06a,Kha07a,Pie09a,Tan10a}, brought a first level of understanding of these states, characterized by sub-threshold Coulomb resonances with very large charging energies (10-50 meV) compared to electrostatically defined quantum dots (usually a few meV). Toward quantum manipulation, it became essential to understand the impact of the transistor environment on the dopant's spectral and coherence properties. The complex three-dimensional (3D) geometry of a transistor, with finite electric fields and the close proximity of dielectric or metallic interfaces, and the sensitivity of the exchange interaction to the precise dopant location in the crystal~\cite{Koi01a} make this task challenging. The dopant's location is {\it a priori} unknown in a transistor since it results either from a stochastic diffusion process from the highly doped source or drain electrodes~\cite{Pie09a}, or from ion implantation with a few nanometers accuracy~\cite{Roc12a}. To address these questions, single-gate transport has evolved with the development of atomistic simulations, multigate devices and dynamical techniques, together with the emergence of single-atom devices specifically designed for quantum computation purposes.\\

The nanometer extent of a dopant's electronic states is suited to a modelling in an atomistic tight-binding (TB) framework where the full spectrum and atomically-resolved wave function spatial densities can be calculated. Such atomistic simulations can be performed over 3D volumes of tens of nanometers corresponding to millions of silicon atoms, and device effects such as electric fields, interfaces, and strain can be incorporated into the same framework. Furthermore, the atomistic TB basis is spin resolved, and captures the effect of spin-orbit and magnetic fields, which have led to the understanding of donor g-factors~\cite{Rah09b} and spin-lattice relaxation times~\cite{Hsu14a}. Effective-mass approaches typically require a multi-valley conduction band basis~\cite{Gam15a}, but they rely on adjustable parameters and lack atomic resolution. Density functional theories typically are size-limited and not amenable to device simulations, despite being atomistic and first-principles~\cite{Swi20a}. However, both of these methods have also advanced recently in accuracy and capability~\cite{Gam15a,Swi20a}.\\

\begin{figure}
\begin{center}
\hspace{-0.05cm}
\includegraphics[width=0.5\textwidth]{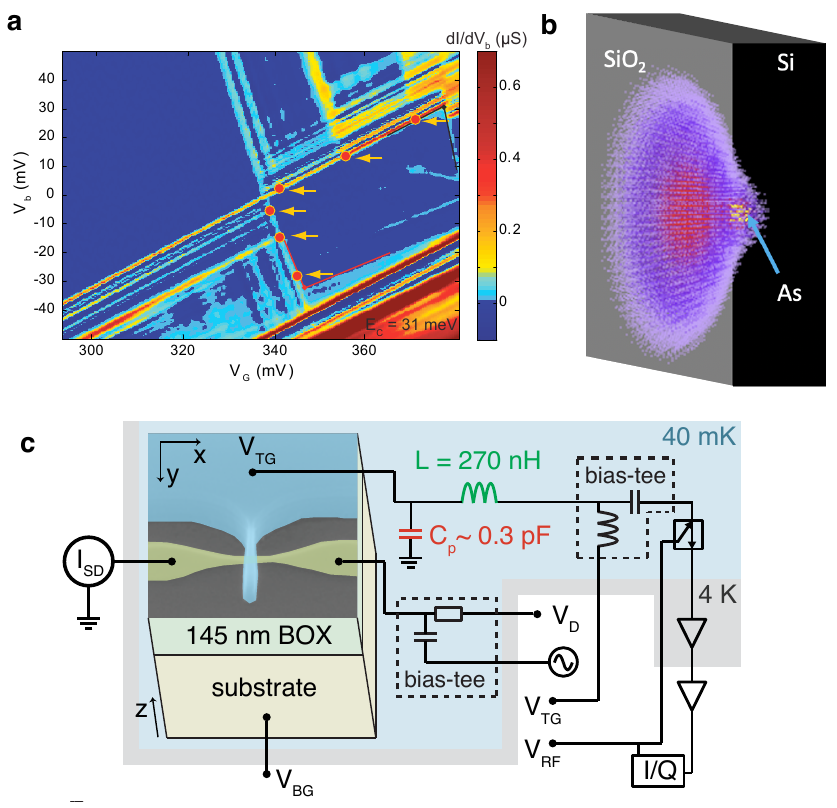}
\caption{\textbf{Single dopant transport, atomistic modelling and reflectometry.} \textbf{a} Stability diagram of the conductance through a FinFET channel is measured at low temperatures versus source-drain ($V_b$) and gate ($V_g$) voltages. The diamond shape is the hallmark of Coulomb blockade. The charging energy of the donor can be directly obtained from the height of the Coulomb diamond, and excited states energies can be identified as conductance resonances within each charge states (yellow arrows)~\cite{Lan08a}. \textbf{b} Three-dimensional representation of the charge distribution $|\psi^2|$ of an electron bound to a single As dopant (located in the black region) close to a gate. The gray region represents the silicon-oxide interface. At finite electric field, the wave function is delocalized between the donor site and the $\rm{Si{/}SiO_2}$ interface~\cite{Lan08a}. \textbf{c} Scanning electron microscope image of a multigate transistor, with the schematic diagram of the experimental setup for DC transport and RF gate reflectometry measurements.~\cite{Hei18a}.}
\label{fig_transistor}
\end{center}
\end{figure}

A joint work between experimental transport and atomistic simulations led to a complete description of a hybrid molecular state that can be formed between a dopant and a quantum dot pinned to the gate interface of a transistor~\cite{Lan08a,Rah09a} (see {\bf Figure~\ref{fig_transistor}}a-b). For the first time, it was possible to use spectral information to pinpoint dopant positions and electric fields whilst providing direct evidence for a Stark shift in dopant systems, essential to Kane's proposal. These simulations have now been extended beyond mean-field and the single electron picture to a full configuration interaction (FCI) framework, where a multielectron wave function can be calculated from all molecular configurations (ground and excited) of the system, each represented by a Slater determinant. These FCI calculations can be used, for instance, to optimize the exchange interaction in multidopant systems~\cite{Wan16a,Voi20a}.\\

Multiple-gate devices have further extended the playground to characterize dopant states in transistors. The substrate of silicon-on-insulator devices can be used as a global back-gate~\cite{Kha07a,Roc12a} to modify the top-gate dopant resonance threshold, and the capacitance coupling matrix obtained from this dependence can be related to the dopant's position. Moreover, changing the gate voltage conditions modifies the electric field obtained from the site, which is reflected in the valley composition of the hybrid dopant/quantum dot state~\cite{Ver13a}. Dual top-gate devices have also been developed to control transport through multiple dopants, which can, for instance, facilitate dopant spectroscopy by filtering the fluctuating density of states of the reservoirs~\cite{Roc12a}.\\

Beyond direct transport, dynamical techniques have also been developed to probe atomic states in transistors. A notable example is the development of gate-based reflectometry, where the quantum capacitance between two dopants or a dopant and a reservoir shifts the
resonance of a tank circuit attached to the gate~\cite{Ver14a}. This technique can detect dopant states on a microsecond time scale instead of the usual millisecond bandwidth of direct transport, and it can also probe dopants which are not sufficiently tunnel coupled to the source or the drain electrode to induce a detectable direct current. Finally, the coherence and coupling of atom systems to their environment can be probed in dynamical experiments using multigate devices. Key highlights include probing a donor's charge coherence time $T_2$ using Landau-Zener microwave irradiations~\cite{Dup13a} or the relaxation time $T_1$ of a coupled acceptor system via timed manipulation (see Figure~\ref{fig_transistor}c)~\cite{Hei18a}.\\

The emergence of single and coupled atom transistors has paved the way for the development of other CMOS-based dopant devices primarily dedicated to quantum information processing, based on precise single ion implantation~\cite{Muh14a} or atomically precise scanning probe lithography~\cite{Fue12a}. Similarly to transistors, multigate devices and fast techniques have led to the demonstration of the  building blocks required for quantum information processing, such as Pauli-spin blockade~\cite{Web14a}, fast detection and single and two-qubit manipulations\cite{Muh14a,He19a}. The atomistic modeling of dopant states, including hyperfine interactions and multielectron charging energies, is an essential ingredient in the metrology of these devices as well~\cite{Web14a}.
 
\section{Quantum State Imaging}
\label{sec:state}
The ideal situation for the design of a quantum computer would be to have complete knowledge of the dopant wave function, since all physically meaningful properties of a quantum system derive from the wave function. Kane's original proposal for donor-based quantum computation\cite{Kan98a} recognized the importance of the donor electron wave function on the operation of the proposed devices. This knowledge, when combined with the dopant positions relative to each other and relative to Si/SiO$_2$ interfaces, single-electron transistors and reservoirs in the device, are inextricably linked to the device operation. It is with this motivation that we describe in this section experiments and theory that provide the most sophisticated level of understanding to date of the bound state electronic wave functions of shallow dopants in silicon, and their interactions with each other and their environment. The challenge of understanding the electronic wave functions of dopants well enough to design a quantum computer is that the wave functions are large enough to be difficult to calculate even using powerful computers, but small enough to be difficult to probe directly in experiments.\\

While many types of experiments have been employed to probe individual dopants\cite{Koe11a}, the type that has revealed the most information about dopant wave functions is low-temperature scanning tunneling microscopy (STM). In the earliest experiments, the cleaved (110) surface that is free of surface states in GaAs played a key role. STM provided the first high resolution signature of the electrostatic potential of a single dopant, a C donor in GaAs\cite{Fee92a}, and the first spatially resolved measurement of the wave function of a single dopant, a Mn acceptor in GaAs\cite{Yak04a}.\\

\begin{figure*}
\begin{center}
\hspace{-0.05cm}
\includegraphics[width=0.9\textwidth]{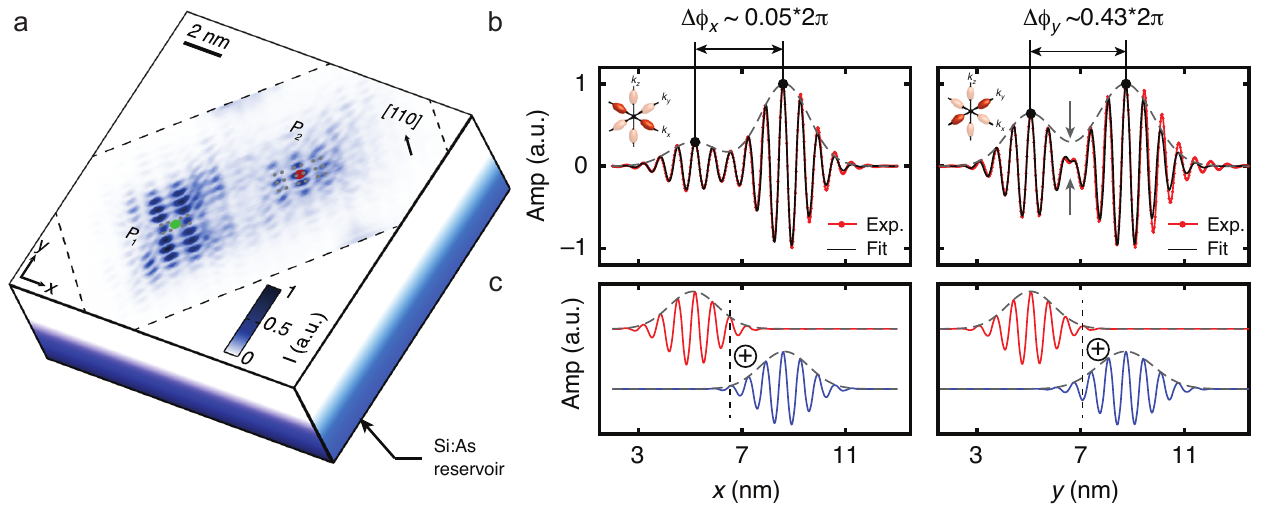}
\caption{\textbf{Measurement of coupled donor wave function~\cite{Voi20a}.} \textbf{a} Real-space single-electron transport current map through the molecular state of two coupled donors labelled P$_1$ and P$_2$. The measurement is a tunneling probability map for the transition between the two-electron state to the one-electron state. \textbf{b} The experimental data (red) can be fit to an analytic model (black) that captures the effective difference in phase between the valleys, the valley frequency, and the anisotropy of the wave functions, which are the key parameters to understand the suppression of exchange due to valley interference. \textbf{c} Evaluation of fitting functions separately for each donor.}
\label{figSTM}
\end{center}
\end{figure*}

Measurement of otherwise well-isolated single dopants and interacting dopants  in silicon was accomplished later, taking advantage of more sophisticated sample and device preparation compared to GaAs. The first innovation was establishing a well-defined double barrier system for resonant electron tunneling. This was first accomplished by rapid vacuum annealing on heavily doped silicon wafers. The annealing creates a depletion region of about 20\,nm between a highly doped reservoir and the silicon surface\cite{Pit12a}, which is terminated by atomic hydrogen. Well-isolated dopants and dopant pairs can be found in this depletion region, which enabled the first observation of the bound states of single donors\cite{Sin14a,Sal14a} and acceptors\cite{Mol15a}, and the first observation of coupled acceptor atoms\cite{Sal16a}, solid-state versions of molecular hydrogen. The essential idea is to perform spatially resolved tunneling spectroscopy with a single bound electronic transition in the resonant tunneling bias window consisting of a scanned tip and the electronic reservoir, all held at 4\,K in a low-temperature STM. Then, the tip is raster scanned to measure the spatial variation of the resonant tunneling current~\cite{Sal14a}. More recent experiments employ more sophisticated fabrication to gain more control over the states being measured, first with P-donors incorporated through PH$_3$ dosing\cite{Sch03a}, and later patterned, quasi two-dimensional reservoirs formed by selective area ion implantation\cite{Ng20a}.\\

This first generation of experiments on single donors enabled the first direct observation of the quantum interference of the valley degree of freedom through standing wave patterns observed in a single donor bound state\cite{Sal14a}. An analysis of these patterns showed that the valley degree of freedom remains relatively unchanged for weak electric fields even when a donor is within 2.5\,nm of the surface\cite{Sal14a}, and enabled parameter extraction for Kohn-Luttinger six-valley wave functions\cite{Sar16a}. More sophisticated analysis of the images was used to pinpoint exact positions of the atoms\cite{Usm16a,Bra17a}. The work on coupled acceptor dopants\cite{Sal16a} yielded an estimate of the strength of the electron-electron correlations, through an analysis of interference processes between the site-centered orbitals. This can be understood by an effective Hubbard interaction parameter $U/t$, where $U$ is the on-site Coulomb repulsion, and $t$ is the interdopant tunneling. This is important for another application of quantum information, the so-called analog quantum simulator\cite{Geo14a}.\\

Newer experiments employed a greater deal of control over the single electron reservoirs and dopant positions to provide a better understanding of qubit interactions. A typical tunnel current image of a coupled-donor molecule is shown in {\bf Figure~\ref{figSTM}}a\cite{Voi20a}. Transformation to Fourier space allows the components of the donor wave function associated with interference between out of plane ($\pm z$) and in-plane ($\pm x$, $\pm y$) valleys to be identified. When these features are transformed back to real space, the standing wave patterns on the donors and their relevant phase difference can be directly visualized (Figure~\ref{figSTM}b-c). This phase difference is relevant because it gives rise to large changes in exchange coupling relevant for two-qubit gates for only atomic-scale displacements in donor position\cite{Koi01a,Wel05a,Wan16a}, which is especially important because exchange-based two-qubit gates have recently been experimentally demonstrated\cite{He19a,Mad20a}. The full two-dimensional analysis, including the observed ellipticity of the orbitals, reveals a fabrication strategy to minimize these variations. The donors can have disorder in their in-plane position, but should occupy the same (100) crystal plane.\\ 

The ability to probe wave functions using STM is currently being extended further to devices fabricated {\it in situ} with atomic-scale dopant placement and electrodes controlled by external voltages. We expect that this will bring new opportunities for measurement and manipulation of multidonor systems. The very first result towards this direction has already been reported, where a tip-induced quantum dot state is controlled by locally ion-implanted electrodes\cite{Ng20a}.

\section{Outlook}
\label{sec:outlook}

In terms of dopants, the main focus in the quantum information science community has been on phosphorus in silicon due to its ability to bind an electron that demonstrates long coherence times. Other dopant and defect sites have emerged in the recent years with the promise of additional functionality. Examples are acceptors \cite{Hei18a,Kob20a}, systems with high-spin nuclei \cite{Asa20a}, and optically active centers \cite{Yin13a,Ber20a}.\\

Acceptors in silicon exhibit strong spin-orbit coupling to enable an all electrical control approach. This had already been identified as advantageous in the early days of quantum computing \cite{Ber05a}. The spin-orbit coupling leads to a pseudo spin 3/2 ground state, and this fourfold degeneracy that can be lifted by strain to separate the two Kramers doublets and enable electrical transitions within these due to mixing. Systems with intrinsic spin-orbit coupling have long been identified as being attractive for quantum information science due to the ability to electrically drive spin transitions without the need for magnetic field gradients, but were challenged due to short coherence times \cite{Pet12a}. Recent progress includes accessing spin dynamics of a singlet-triplet transition of two tunnel coupled acceptors using reflectometry \cite{Hei18a}, and the demonstration that acceptor pseudo spins in strained $^{28}$Si can achieve similar coherence times to P:Si \cite{Kob20a}. Proposals further investigate how to improve the robustness against charge noise and to couple them to microwave photons for long-distance coupling \cite{Sal16b} to make acceptor dopants in silicon an interesting material system for quantum information science.\\

Electrical manipulation was also recently extended to nuclear spins, using a 7/2 nuclear spin in a single antimony $^{123}$Sb donor in strained silicon \cite{Asa20a}, where a resonant electric field induces nuclear spin transitions by modulating the nuclear quadrupole interaction. Besides the appealing ability to electrically control the nuclear spin to avoid the need for on-chip oscillating magnetic fields, this system also allows the study of quantum chaotic dynamics in a single particle \cite{Mou18a} and links to digital quantum simulations \cite{Sie19a}. Dynamical strain based on mechanical resonators is an interesting direction for this system as well as for acceptors since it may open a pathway to coherent spin-phonon coupling \cite{Kur15a,Rus13a}. Quantum state imaging is a valuable tool for these systems since they derive their functionality from the interaction between the quantum system and perturbation in the environment \cite{Mol13a}.\\

Long-distance communication between quantum systems via photons at telecommunication wavelengths is a major outstanding goal of experimental quantum science. Silicon is not an attractive host for optical applications due to the indirect bandgap. However, dopant and defect sites in silicon can present transitions within this bandwidth. Rare-earth ions such as erbium in silicon have an inner shell (i.e., not related to the silicon band structure), with optical transitions in the telecom window. The potential for long spin coherence times \cite{Zho15a} sparked ensemble \cite{Wei20pp} and single site experiments \cite{Yin13a} challenged by the wide selection of site symmetries in this host. Recently, defects known as radiation damage centers, such as the so called T-center, have been investigated due to their robustness, emission around 1.35\,$\mu$m, and long spin coherence times that have been studied in isotopically enriched silicon \cite{Ber20a}.

\section{Conclusion}
\label{sec:conclusion}

Dopant atoms and defects in semiconductors offer a complete range of quantum systems with exceptional coherence times due to their tightly bound wave functions ranging from angstroms to a few nanometers. Experiments performed over the last decade have established these single-atom quantum systems as promising building blocks of quantum computers, sensors and memories. While probing single-atoms within an environment of millions of host atoms was a major scientific challenge 20 years ago, advancement in measurement, fabrication, and simulation techniques have now enabled the routine detection and manipulation of their quantum states. Electrical transport and multigate control can probe the electrostatic coupling of single dopants with their semiconductor environment, while quantum state imaging by scanning tunneling spectroscopy can provide an atomically resolved probe of the bound wave functions. Atomistic simulations can provide an independent picture of the dopant-environment interactions at the level of wave functions. Together, these techniques have helped to advance our knowledge of single-dopant quantum electronics and geared up this community for scaled-up quantum technologies. 

\section{Acknowledgements}
This review article is dedicated to Marc Sanquer. We acknowledge support from the ARC Centre of Excellence for Quantum Computation and Communication Technology (CE170100012), Silicon Quantum Computing Pty Ltd., and from the U.S. Army Research Office (W911NF-17-1-0202). J.S. acknowledges support from the National Science and Engineering Research Council of Canada.

\section{Conflict of Interest}
On behalf of all authors, the corresponding author states that there is no conflict of interest.

\bibliographystyle{aipnum4-1.bst}
\bibliography{MRS_donor_fix}

\begin{thebibliography}{62}%
\makeatletter
\providecommand \@ifxundefined [1]{%
 \@ifx{#1\undefined}
}%
\providecommand \@ifnum [1]{%
 \ifnum #1\expandafter \@firstoftwo
 \else \expandafter \@secondoftwo
 \fi
}%
\providecommand \@ifx [1]{%
 \ifx #1\expandafter \@firstoftwo
 \else \expandafter \@secondoftwo
 \fi
}%
\providecommand \natexlab [1]{#1}%
\providecommand \enquote  [1]{``#1''}%
\providecommand \bibnamefont  [1]{#1}%
\providecommand \bibfnamefont [1]{#1}%
\providecommand \citenamefont [1]{#1}%
\providecommand \href@noop [0]{\@secondoftwo}%
\providecommand \href [0]{\begingroup \@sanitize@url \@href}%
\providecommand \@href[1]{\@@startlink{#1}\@@href}%
\providecommand \@@href[1]{\endgroup#1\@@endlink}%
\providecommand \@sanitize@url [0]{\catcode `\\12\catcode `\$12\catcode
  `\&12\catcode `\#12\catcode `\^12\catcode `\_12\catcode `\%12\relax}%
\providecommand \@@startlink[1]{}%
\providecommand \@@endlink[0]{}%
\providecommand \url  [0]{\begingroup\@sanitize@url \@url }%
\providecommand \@url [1]{\endgroup\@href {#1}{\urlprefix }}%
\providecommand \urlprefix  [0]{URL }%
\providecommand \Eprint [0]{\href }%
\providecommand \doibase [0]{http://dx.doi.org/}%
\providecommand \selectlanguage [0]{\@gobble}%
\providecommand \bibinfo  [0]{\@secondoftwo}%
\providecommand \bibfield  [0]{\@secondoftwo}%
\providecommand \translation [1]{[#1]}%
\providecommand \BibitemOpen [0]{}%
\providecommand \bibitemStop [0]{}%
\providecommand \bibitemNoStop [0]{.\EOS\space}%
\providecommand \EOS [0]{\spacefactor3000\relax}%
\providecommand \BibitemShut  [1]{\csname bibitem#1\endcsname}%
\let\auto@bib@innerbib\@empty
\bibitem [{\citenamefont {Feynman}(1982)}]{Fey82a}%
  \BibitemOpen
  \bibfield  {author} {\bibinfo {author} {\bibfnamefont {R.~P.}\ \bibnamefont
  {Feynman}},\ }\href {\doibase 10.1007/BF02650179} {\bibfield  {journal}
  {\bibinfo  {journal} {International Journal of Theoretical Physics}\ }\textbf
  {\bibinfo {volume} {21}},\ \bibinfo {pages} {467} (\bibinfo {year}
  {1982})}\BibitemShut {NoStop}%
\bibitem [{\citenamefont {Cirac}\ and\ \citenamefont {Zoller}(1995)}]{Cir95a}%
  \BibitemOpen
  \bibfield  {author} {\bibinfo {author} {\bibfnamefont {J.~I.}\ \bibnamefont
  {Cirac}}\ and\ \bibinfo {author} {\bibfnamefont {P.}~\bibnamefont {Zoller}},\
  }\href {\doibase 10.1103/PhysRevLett.74.4091} {\bibfield  {journal} {\bibinfo
   {journal} {Phys. Rev. Lett.}\ }\textbf {\bibinfo {volume} {74}},\ \bibinfo
  {pages} {4091} (\bibinfo {year} {1995})}\BibitemShut {NoStop}%
\bibitem [{\citenamefont {Blais}\ \emph {et~al.}(2004)\citenamefont {Blais},
  \citenamefont {Huang}, \citenamefont {Wallraff}, \citenamefont {Girvin},\
  and\ \citenamefont {Schoelkopf}}]{Bla04a}%
  \BibitemOpen
  \bibfield  {author} {\bibinfo {author} {\bibfnamefont {A.}~\bibnamefont
  {Blais}}, \bibinfo {author} {\bibfnamefont {R.-S.}\ \bibnamefont {Huang}},
  \bibinfo {author} {\bibfnamefont {A.}~\bibnamefont {Wallraff}}, \bibinfo
  {author} {\bibfnamefont {S.~M.}\ \bibnamefont {Girvin}}, \ and\ \bibinfo
  {author} {\bibfnamefont {R.~J.}\ \bibnamefont {Schoelkopf}},\ }\href
  {\doibase 10.1103/PhysRevA.69.062320} {\bibfield  {journal} {\bibinfo
  {journal} {Phys. Rev. A}\ }\textbf {\bibinfo {volume} {69}},\ \bibinfo
  {pages} {062320} (\bibinfo {year} {2004})}\BibitemShut {NoStop}%
\bibitem [{\citenamefont {Loss}\ and\ \citenamefont
  {DiVincenzo}(1998)}]{Los98a}%
  \BibitemOpen
  \bibfield  {author} {\bibinfo {author} {\bibfnamefont {D.}~\bibnamefont
  {Loss}}\ and\ \bibinfo {author} {\bibfnamefont {D.~P.}\ \bibnamefont
  {DiVincenzo}},\ }\href {\doibase 10.1103/PhysRevA.57.120} {\bibfield
  {journal} {\bibinfo  {journal} {Phys. Rev. A}\ }\textbf {\bibinfo {volume}
  {57}},\ \bibinfo {pages} {120} (\bibinfo {year} {1998})}\BibitemShut
  {NoStop}%
\bibitem [{\citenamefont {Kane}(1998)}]{Kan98a}%
  \BibitemOpen
  \bibfield  {author} {\bibinfo {author} {\bibfnamefont {B.~E.}\ \bibnamefont
  {Kane}},\ }\href {\doibase 10.1038/30156} {\bibfield  {journal} {\bibinfo
  {journal} {Nature}\ }\textbf {\bibinfo {volume} {393}},\ \bibinfo {pages}
  {133} (\bibinfo {year} {1998})}\BibitemShut {NoStop}%
\bibitem [{\citenamefont {DiVincenzo}(2000)}]{DiV00a}%
  \BibitemOpen
  \bibfield  {author} {\bibinfo {author} {\bibfnamefont {D.~P.}\ \bibnamefont
  {DiVincenzo}},\ }\href {\doibase
  https://doi.org/10.1002/1521-3978(200009)48:9/11<771::AID-PROP771>3.0.CO;2-E}
  {\bibfield  {journal} {\bibinfo  {journal} {Fortschritte der Physik}\
  }\textbf {\bibinfo {volume} {48}},\ \bibinfo {pages} {771} (\bibinfo {year}
  {2000})}\BibitemShut {NoStop}%
\bibitem [{\citenamefont {Watson}\ \emph {et~al.}(2018)\citenamefont {Watson},
  \citenamefont {Philips}, \citenamefont {Kawakami}, \citenamefont {Ward},
  \citenamefont {Scarlino}, \citenamefont {Veldhorst}, \citenamefont {Savage},
  \citenamefont {Lagally}, \citenamefont {Friesen}, \citenamefont
  {Coppersmith}, \citenamefont {Eriksson},\ and\ \citenamefont
  {Vandersypen}}]{Wat18a}%
  \BibitemOpen
  \bibfield  {author} {\bibinfo {author} {\bibfnamefont {T.~F.}\ \bibnamefont
  {Watson}}, \bibinfo {author} {\bibfnamefont {S.~G.~J.}\ \bibnamefont
  {Philips}}, \bibinfo {author} {\bibfnamefont {E.}~\bibnamefont {Kawakami}},
  \bibinfo {author} {\bibfnamefont {D.~R.}\ \bibnamefont {Ward}}, \bibinfo
  {author} {\bibfnamefont {P.}~\bibnamefont {Scarlino}}, \bibinfo {author}
  {\bibfnamefont {M.}~\bibnamefont {Veldhorst}}, \bibinfo {author}
  {\bibfnamefont {D.~E.}\ \bibnamefont {Savage}}, \bibinfo {author}
  {\bibfnamefont {M.~G.}\ \bibnamefont {Lagally}}, \bibinfo {author}
  {\bibfnamefont {M.}~\bibnamefont {Friesen}}, \bibinfo {author} {\bibfnamefont
  {S.~N.}\ \bibnamefont {Coppersmith}}, \bibinfo {author} {\bibfnamefont
  {M.~A.}\ \bibnamefont {Eriksson}}, \ and\ \bibinfo {author} {\bibfnamefont
  {L.~M.~K.}\ \bibnamefont {Vandersypen}},\ }\href
  {http://dx.doi.org/10.1038/nature25766} {\bibfield  {journal} {\bibinfo
  {journal} {Nature}\ }\textbf {\bibinfo {volume} {555}},\ \bibinfo {pages}
  {633} (\bibinfo {year} {2018})}\BibitemShut {NoStop}%
\bibitem [{\citenamefont {Veldhorst}\ \emph {et~al.}(2015)\citenamefont
  {Veldhorst}, \citenamefont {Yang}, \citenamefont {Hwang}, \citenamefont
  {Huang}, \citenamefont {Dehollain}, \citenamefont {Muhonen}, \citenamefont
  {Simmons}, \citenamefont {Laucht}, \citenamefont {Hudson}, \citenamefont
  {Itoh}, \citenamefont {Morello},\ and\ \citenamefont {Dzurak}}]{Vel15a}%
  \BibitemOpen
  \bibfield  {author} {\bibinfo {author} {\bibfnamefont {M.}~\bibnamefont
  {Veldhorst}}, \bibinfo {author} {\bibfnamefont {C.~H.}\ \bibnamefont {Yang}},
  \bibinfo {author} {\bibfnamefont {J.~C.~C.}\ \bibnamefont {Hwang}}, \bibinfo
  {author} {\bibfnamefont {W.}~\bibnamefont {Huang}}, \bibinfo {author}
  {\bibfnamefont {J.~P.}\ \bibnamefont {Dehollain}}, \bibinfo {author}
  {\bibfnamefont {J.~T.}\ \bibnamefont {Muhonen}}, \bibinfo {author}
  {\bibfnamefont {S.}~\bibnamefont {Simmons}}, \bibinfo {author} {\bibfnamefont
  {A.}~\bibnamefont {Laucht}}, \bibinfo {author} {\bibfnamefont {F.~E.}\
  \bibnamefont {Hudson}}, \bibinfo {author} {\bibfnamefont {K.~M.}\
  \bibnamefont {Itoh}}, \bibinfo {author} {\bibfnamefont {A.}~\bibnamefont
  {Morello}}, \ and\ \bibinfo {author} {\bibfnamefont {A.~S.}\ \bibnamefont
  {Dzurak}},\ }\href {http://dx.doi.org/10.1038/nature15263} {\bibfield
  {journal} {\bibinfo  {journal} {Nature}\ }\textbf {\bibinfo {volume} {526}},\
  \bibinfo {pages} {410} (\bibinfo {year} {2015})}\BibitemShut {NoStop}%
\bibitem [{\citenamefont {Muhonen}\ \emph {et~al.}(2014)\citenamefont
  {Muhonen}, \citenamefont {Dehollain}, \citenamefont {Laucht}, \citenamefont
  {Hudson}, \citenamefont {Kalra}, \citenamefont {Sekiguchi}, \citenamefont
  {Itoh}, \citenamefont {Jamieson}, \citenamefont {McCallum}, \citenamefont
  {Dzurak},\ and\ \citenamefont {Morello}}]{Muh14a}%
  \BibitemOpen
  \bibfield  {author} {\bibinfo {author} {\bibfnamefont {J.~T.}\ \bibnamefont
  {Muhonen}}, \bibinfo {author} {\bibfnamefont {J.~P.}\ \bibnamefont
  {Dehollain}}, \bibinfo {author} {\bibfnamefont {A.}~\bibnamefont {Laucht}},
  \bibinfo {author} {\bibfnamefont {F.~E.}\ \bibnamefont {Hudson}}, \bibinfo
  {author} {\bibfnamefont {R.}~\bibnamefont {Kalra}}, \bibinfo {author}
  {\bibfnamefont {T.}~\bibnamefont {Sekiguchi}}, \bibinfo {author}
  {\bibfnamefont {K.~M.}\ \bibnamefont {Itoh}}, \bibinfo {author}
  {\bibfnamefont {D.~N.}\ \bibnamefont {Jamieson}}, \bibinfo {author}
  {\bibfnamefont {J.~C.}\ \bibnamefont {McCallum}}, \bibinfo {author}
  {\bibfnamefont {A.~S.}\ \bibnamefont {Dzurak}}, \ and\ \bibinfo {author}
  {\bibfnamefont {A.}~\bibnamefont {Morello}},\ }\href
  {https://doi.org/10.1038/nnano.2014.211} {\bibfield  {journal} {\bibinfo
  {journal} {Nature Nanotechnology}\ }\textbf {\bibinfo {volume} {9}},\
  \bibinfo {pages} {986} (\bibinfo {year} {2014})}\BibitemShut {NoStop}%
\bibitem [{\citenamefont {He}\ \emph {et~al.}(2019)\citenamefont {He},
  \citenamefont {Gorman}, \citenamefont {Keith}, \citenamefont {Kranz},
  \citenamefont {Keizer},\ and\ \citenamefont {Simmons}}]{He19a}%
  \BibitemOpen
  \bibfield  {author} {\bibinfo {author} {\bibfnamefont {Y.}~\bibnamefont
  {He}}, \bibinfo {author} {\bibfnamefont {S.~K.}\ \bibnamefont {Gorman}},
  \bibinfo {author} {\bibfnamefont {D.}~\bibnamefont {Keith}}, \bibinfo
  {author} {\bibfnamefont {L.}~\bibnamefont {Kranz}}, \bibinfo {author}
  {\bibfnamefont {J.~G.}\ \bibnamefont {Keizer}}, \ and\ \bibinfo {author}
  {\bibfnamefont {M.~Y.}\ \bibnamefont {Simmons}},\ }\href
  {https://doi.org/10.1038/s41586-019-1381-2} {\bibfield  {journal} {\bibinfo
  {journal} {Nature}\ }\textbf {\bibinfo {volume} {571}},\ \bibinfo {pages}
  {371} (\bibinfo {year} {2019})}\BibitemShut {NoStop}%
\bibitem [{\citenamefont {Dellow}\ \emph {et~al.}(1992)\citenamefont {Dellow},
  \citenamefont {Beton}, \citenamefont {Langerak}, \citenamefont {Foster},
  \citenamefont {Main}, \citenamefont {Eaves}, \citenamefont {Henini},
  \citenamefont {Beaumont},\ and\ \citenamefont {Wilkinson}}]{Del92a}%
  \BibitemOpen
  \bibfield  {author} {\bibinfo {author} {\bibfnamefont {M.~W.}\ \bibnamefont
  {Dellow}}, \bibinfo {author} {\bibfnamefont {P.~H.}\ \bibnamefont {Beton}},
  \bibinfo {author} {\bibfnamefont {C.~J. G.~M.}\ \bibnamefont {Langerak}},
  \bibinfo {author} {\bibfnamefont {T.~J.}\ \bibnamefont {Foster}}, \bibinfo
  {author} {\bibfnamefont {P.~C.}\ \bibnamefont {Main}}, \bibinfo {author}
  {\bibfnamefont {L.}~\bibnamefont {Eaves}}, \bibinfo {author} {\bibfnamefont
  {M.}~\bibnamefont {Henini}}, \bibinfo {author} {\bibfnamefont {S.~P.}\
  \bibnamefont {Beaumont}}, \ and\ \bibinfo {author} {\bibfnamefont {C.~D.~W.}\
  \bibnamefont {Wilkinson}},\ }\href {\doibase 10.1103/PhysRevLett.68.1754}
  {\bibfield  {journal} {\bibinfo  {journal} {Phys. Rev. Lett.}\ }\textbf
  {\bibinfo {volume} {68}},\ \bibinfo {pages} {1754} (\bibinfo {year}
  {1992})}\BibitemShut {NoStop}%
\bibitem [{\citenamefont {Asenov}\ \emph {et~al.}(2003)\citenamefont {Asenov},
  \citenamefont {Brown}, \citenamefont {Davies}, \citenamefont {Kaya},\ and\
  \citenamefont {Slavcheva}}]{Ase03a}%
  \BibitemOpen
  \bibfield  {author} {\bibinfo {author} {\bibfnamefont {A.}~\bibnamefont
  {Asenov}}, \bibinfo {author} {\bibfnamefont {A.}~\bibnamefont {Brown}},
  \bibinfo {author} {\bibfnamefont {J.}~\bibnamefont {Davies}}, \bibinfo
  {author} {\bibfnamefont {S.}~\bibnamefont {Kaya}}, \ and\ \bibinfo {author}
  {\bibfnamefont {G.}~\bibnamefont {Slavcheva}},\ }\href {\doibase
  10.1109/TED.2003.815862} {\bibfield  {journal} {\bibinfo  {journal} {IEEE
  Transactions on Electron Devices}\ }\textbf {\bibinfo {volume} {50}},\
  \bibinfo {pages} {1837} (\bibinfo {year} {2003})}\BibitemShut {NoStop}%
\bibitem [{\citenamefont {Sellier}\ \emph {et~al.}(2006)\citenamefont
  {Sellier}, \citenamefont {Lansbergen}, \citenamefont {Caro}, \citenamefont
  {Rogge}, \citenamefont {Collaert}, \citenamefont {Ferain}, \citenamefont
  {Jurczak},\ and\ \citenamefont {Biesemans}}]{Sel06a}%
  \BibitemOpen
  \bibfield  {author} {\bibinfo {author} {\bibfnamefont {H.}~\bibnamefont
  {Sellier}}, \bibinfo {author} {\bibfnamefont {G.~P.}\ \bibnamefont
  {Lansbergen}}, \bibinfo {author} {\bibfnamefont {J.}~\bibnamefont {Caro}},
  \bibinfo {author} {\bibfnamefont {S.}~\bibnamefont {Rogge}}, \bibinfo
  {author} {\bibfnamefont {N.}~\bibnamefont {Collaert}}, \bibinfo {author}
  {\bibfnamefont {I.}~\bibnamefont {Ferain}}, \bibinfo {author} {\bibfnamefont
  {M.}~\bibnamefont {Jurczak}}, \ and\ \bibinfo {author} {\bibfnamefont
  {S.}~\bibnamefont {Biesemans}},\ }\href {\doibase
  10.1103/PhysRevLett.97.206805} {\bibfield  {journal} {\bibinfo  {journal}
  {Phys. Rev. Lett.}\ }\textbf {\bibinfo {volume} {97}},\ \bibinfo {pages}
  {206805} (\bibinfo {year} {2006})}\BibitemShut {NoStop}%
\bibitem [{\citenamefont {Khalafalla}\ \emph {et~al.}(2007)\citenamefont
  {Khalafalla}, \citenamefont {Ono}, \citenamefont {Nishiguchi},\ and\
  \citenamefont {Fujiwara}}]{Kha07a}%
  \BibitemOpen
  \bibfield  {author} {\bibinfo {author} {\bibfnamefont {M.~A.~H.}\
  \bibnamefont {Khalafalla}}, \bibinfo {author} {\bibfnamefont
  {Y.}~\bibnamefont {Ono}}, \bibinfo {author} {\bibfnamefont {K.}~\bibnamefont
  {Nishiguchi}}, \ and\ \bibinfo {author} {\bibfnamefont {A.}~\bibnamefont
  {Fujiwara}},\ }\href {\doibase https://doi.org/10.1063/1.2824579} {\bibfield
  {journal} {\bibinfo  {journal} {Applied Physics Letters}\ }\textbf {\bibinfo
  {volume} {91}},\ \bibinfo {pages} {263513} (\bibinfo {year}
  {2007})}\BibitemShut {NoStop}%
\bibitem [{\citenamefont {Pierre}\ \emph {et~al.}(2009)\citenamefont {Pierre},
  \citenamefont {Wacquez}, \citenamefont {Jehl}, \citenamefont {Sanquer},
  \citenamefont {Vinet},\ and\ \citenamefont {Cueto}}]{Pie09a}%
  \BibitemOpen
  \bibfield  {author} {\bibinfo {author} {\bibfnamefont {M.}~\bibnamefont
  {Pierre}}, \bibinfo {author} {\bibfnamefont {R.}~\bibnamefont {Wacquez}},
  \bibinfo {author} {\bibfnamefont {X.}~\bibnamefont {Jehl}}, \bibinfo {author}
  {\bibfnamefont {M.}~\bibnamefont {Sanquer}}, \bibinfo {author} {\bibfnamefont
  {M.}~\bibnamefont {Vinet}}, \ and\ \bibinfo {author} {\bibfnamefont
  {O.}~\bibnamefont {Cueto}},\ }\href {\doibase
  https://doi.org/10.1038/nnano.2009.373} {\bibfield  {journal} {\bibinfo
  {journal} {Nature Nanotechnology}\ }\textbf {\bibinfo {volume} {5}},\
  \bibinfo {pages} {133} (\bibinfo {year} {2009})}\BibitemShut {NoStop}%
\bibitem [{\citenamefont {Tan}\ \emph {et~al.}(2010)\citenamefont {Tan},
  \citenamefont {Chan}, \citenamefont {M{\"o}tt{\"o}nen}, \citenamefont
  {Morello}, \citenamefont {Yang}, \citenamefont {Donkelaar}, \citenamefont
  {Alves}, \citenamefont {Pirkkalainen}, \citenamefont {Jamieson},
  \citenamefont {Clark},\ and\ \citenamefont {Dzurak}}]{Tan10a}%
  \BibitemOpen
  \bibfield  {author} {\bibinfo {author} {\bibfnamefont {K.~Y.}\ \bibnamefont
  {Tan}}, \bibinfo {author} {\bibfnamefont {K.~W.}\ \bibnamefont {Chan}},
  \bibinfo {author} {\bibfnamefont {M.}~\bibnamefont {M{\"o}tt{\"o}nen}},
  \bibinfo {author} {\bibfnamefont {A.}~\bibnamefont {Morello}}, \bibinfo
  {author} {\bibfnamefont {C.}~\bibnamefont {Yang}}, \bibinfo {author}
  {\bibfnamefont {J.~v.}\ \bibnamefont {Donkelaar}}, \bibinfo {author}
  {\bibfnamefont {A.}~\bibnamefont {Alves}}, \bibinfo {author} {\bibfnamefont
  {J.-M.}\ \bibnamefont {Pirkkalainen}}, \bibinfo {author} {\bibfnamefont
  {D.~N.}\ \bibnamefont {Jamieson}}, \bibinfo {author} {\bibfnamefont {R.~G.}\
  \bibnamefont {Clark}}, \ and\ \bibinfo {author} {\bibfnamefont {A.~S.}\
  \bibnamefont {Dzurak}},\ }\href {\doibase https://doi.org/10.1021/nl901635j}
  {\bibfield  {journal} {\bibinfo  {journal} {Nano Letters}\ }\textbf {\bibinfo
  {volume} {10}},\ \bibinfo {pages} {11} (\bibinfo {year} {2010})}\BibitemShut
  {NoStop}%
\bibitem [{\citenamefont {Lansbergen}\ \emph {et~al.}(2008)\citenamefont
  {Lansbergen}, \citenamefont {Rahman}, \citenamefont {Wellard}, \citenamefont
  {Woo}, \citenamefont {Caro}, \citenamefont {Collaert}, \citenamefont
  {Biesemans}, \citenamefont {Klimeck}, \citenamefont {Hollenberg},\ and\
  \citenamefont {Rogge}}]{Lan08a}%
  \BibitemOpen
  \bibfield  {author} {\bibinfo {author} {\bibfnamefont {G.}~\bibnamefont
  {Lansbergen}}, \bibinfo {author} {\bibfnamefont {R.}~\bibnamefont {Rahman}},
  \bibinfo {author} {\bibfnamefont {C.}~\bibnamefont {Wellard}}, \bibinfo
  {author} {\bibfnamefont {I.}~\bibnamefont {Woo}}, \bibinfo {author}
  {\bibfnamefont {J.}~\bibnamefont {Caro}}, \bibinfo {author} {\bibfnamefont
  {N.}~\bibnamefont {Collaert}}, \bibinfo {author} {\bibfnamefont
  {S.}~\bibnamefont {Biesemans}}, \bibinfo {author} {\bibfnamefont
  {G.}~\bibnamefont {Klimeck}}, \bibinfo {author} {\bibfnamefont
  {L.}~\bibnamefont {Hollenberg}}, \ and\ \bibinfo {author} {\bibfnamefont
  {S.}~\bibnamefont {Rogge}},\ }\href {\doibase
  https://doi.org/10.1038/nphys994} {\bibfield  {journal} {\bibinfo  {journal}
  {Nature Physics}\ }\textbf {\bibinfo {volume} {4}},\ \bibinfo {pages} {656}
  (\bibinfo {year} {2008})}\BibitemShut {NoStop}%
\bibitem [{\citenamefont {Salfi}\ \emph {et~al.}(2014)\citenamefont {Salfi},
  \citenamefont {Mol}, \citenamefont {Rahman}, \citenamefont {Klimeck},
  \citenamefont {Simmons}, \citenamefont {Hollenberg},\ and\ \citenamefont
  {Rogge}}]{Sal14a}%
  \BibitemOpen
  \bibfield  {author} {\bibinfo {author} {\bibfnamefont {J.}~\bibnamefont
  {Salfi}}, \bibinfo {author} {\bibfnamefont {J.~A.}\ \bibnamefont {Mol}},
  \bibinfo {author} {\bibfnamefont {R.}~\bibnamefont {Rahman}}, \bibinfo
  {author} {\bibfnamefont {G.}~\bibnamefont {Klimeck}}, \bibinfo {author}
  {\bibfnamefont {M.~Y.}\ \bibnamefont {Simmons}}, \bibinfo {author}
  {\bibfnamefont {L.~C.~L.}\ \bibnamefont {Hollenberg}}, \ and\ \bibinfo
  {author} {\bibfnamefont {S.}~\bibnamefont {Rogge}},\ }\href {\doibase
  https://doi.org/10.1038/nmat3941} {\bibfield  {journal} {\bibinfo  {journal}
  {Nature Materials}\ }\textbf {\bibinfo {volume} {13}},\ \bibinfo {pages}
  {605} (\bibinfo {year} {2014})}\BibitemShut {NoStop}%
\bibitem [{\citenamefont {Koiller}, \citenamefont {Hu},\ and\ \citenamefont
  {Sarma}(2001)}]{Koi01a}%
  \BibitemOpen
  \bibfield  {author} {\bibinfo {author} {\bibfnamefont {B.}~\bibnamefont
  {Koiller}}, \bibinfo {author} {\bibfnamefont {X.}~\bibnamefont {Hu}}, \ and\
  \bibinfo {author} {\bibfnamefont {S.~D.}\ \bibnamefont {Sarma}},\ }\href
  {\doibase 10.1103/PhysRevLett.88.027903} {\bibfield  {journal} {\bibinfo
  {journal} {Phys. Rev. Lett.}\ }\textbf {\bibinfo {volume} {88}},\ \bibinfo
  {pages} {027903} (\bibinfo {year} {2001})}\BibitemShut {NoStop}%
\bibitem [{\citenamefont {Roche}\ \emph {et~al.}(2012)\citenamefont {Roche},
  \citenamefont {Dupont-Ferrier}, \citenamefont {Voisin}, \citenamefont
  {Cobian}, \citenamefont {Jehl}, \citenamefont {Wacquez}, \citenamefont
  {Vinet}, \citenamefont {Niquet},\ and\ \citenamefont {Sanquer}}]{Roc12a}%
  \BibitemOpen
  \bibfield  {author} {\bibinfo {author} {\bibfnamefont {B.}~\bibnamefont
  {Roche}}, \bibinfo {author} {\bibfnamefont {E.}~\bibnamefont
  {Dupont-Ferrier}}, \bibinfo {author} {\bibfnamefont {B.}~\bibnamefont
  {Voisin}}, \bibinfo {author} {\bibfnamefont {M.}~\bibnamefont {Cobian}},
  \bibinfo {author} {\bibfnamefont {X.}~\bibnamefont {Jehl}}, \bibinfo {author}
  {\bibfnamefont {R.}~\bibnamefont {Wacquez}}, \bibinfo {author} {\bibfnamefont
  {M.}~\bibnamefont {Vinet}}, \bibinfo {author} {\bibfnamefont {Y.-M.}\
  \bibnamefont {Niquet}}, \ and\ \bibinfo {author} {\bibfnamefont
  {M.}~\bibnamefont {Sanquer}},\ }\href {\doibase
  10.1103/PhysRevLett.108.206812} {\bibfield  {journal} {\bibinfo  {journal}
  {Phys. Rev. Lett.}\ }\textbf {\bibinfo {volume} {108}},\ \bibinfo {pages}
  {206812} (\bibinfo {year} {2012})}\BibitemShut {NoStop}%
\bibitem [{\citenamefont {Rahman}\ \emph
  {et~al.}(2009{\natexlab{a}})\citenamefont {Rahman}, \citenamefont {Park},
  \citenamefont {Boykin}, \citenamefont {Klimeck}, \citenamefont {Rogge},\ and\
  \citenamefont {Hollenberg}}]{Rah09b}%
  \BibitemOpen
  \bibfield  {author} {\bibinfo {author} {\bibfnamefont {R.}~\bibnamefont
  {Rahman}}, \bibinfo {author} {\bibfnamefont {S.~H.}\ \bibnamefont {Park}},
  \bibinfo {author} {\bibfnamefont {T.~B.}\ \bibnamefont {Boykin}}, \bibinfo
  {author} {\bibfnamefont {G.}~\bibnamefont {Klimeck}}, \bibinfo {author}
  {\bibfnamefont {S.}~\bibnamefont {Rogge}}, \ and\ \bibinfo {author}
  {\bibfnamefont {L.~C.~L.}\ \bibnamefont {Hollenberg}},\ }\href {\doibase
  10.1103/PhysRevB.80.155301} {\bibfield  {journal} {\bibinfo  {journal} {Phys.
  Rev. B}\ }\textbf {\bibinfo {volume} {80}},\ \bibinfo {pages} {155301}
  (\bibinfo {year} {2009}{\natexlab{a}})}\BibitemShut {NoStop}%
\bibitem [{\citenamefont {Hsueh}\ \emph {et~al.}(2014)\citenamefont {Hsueh},
  \citenamefont {B\"uch}, \citenamefont {Tan}, \citenamefont {Wang},
  \citenamefont {Hollenberg}, \citenamefont {Klimeck}, \citenamefont
  {Simmons},\ and\ \citenamefont {Rahman}}]{Hsu14a}%
  \BibitemOpen
  \bibfield  {author} {\bibinfo {author} {\bibfnamefont {Y.-L.}\ \bibnamefont
  {Hsueh}}, \bibinfo {author} {\bibfnamefont {H.}~\bibnamefont {B\"uch}},
  \bibinfo {author} {\bibfnamefont {Y.}~\bibnamefont {Tan}}, \bibinfo {author}
  {\bibfnamefont {Y.}~\bibnamefont {Wang}}, \bibinfo {author} {\bibfnamefont
  {L.~C.~L.}\ \bibnamefont {Hollenberg}}, \bibinfo {author} {\bibfnamefont
  {G.}~\bibnamefont {Klimeck}}, \bibinfo {author} {\bibfnamefont {M.~Y.}\
  \bibnamefont {Simmons}}, \ and\ \bibinfo {author} {\bibfnamefont
  {R.}~\bibnamefont {Rahman}},\ }\href {\doibase
  10.1103/PhysRevLett.113.246406} {\bibfield  {journal} {\bibinfo  {journal}
  {Phys. Rev. Lett.}\ }\textbf {\bibinfo {volume} {113}},\ \bibinfo {pages}
  {246406} (\bibinfo {year} {2014})}\BibitemShut {NoStop}%
\bibitem [{\citenamefont {Gamble}\ \emph {et~al.}(2015)\citenamefont {Gamble},
  \citenamefont {Jacobson}, \citenamefont {Nielsen}, \citenamefont {Baczewski},
  \citenamefont {Moussa}, \citenamefont {Monta\~no},\ and\ \citenamefont
  {Muller}}]{Gam15a}%
  \BibitemOpen
  \bibfield  {author} {\bibinfo {author} {\bibfnamefont {J.~K.}\ \bibnamefont
  {Gamble}}, \bibinfo {author} {\bibfnamefont {N.~T.}\ \bibnamefont
  {Jacobson}}, \bibinfo {author} {\bibfnamefont {E.}~\bibnamefont {Nielsen}},
  \bibinfo {author} {\bibfnamefont {A.~D.}\ \bibnamefont {Baczewski}}, \bibinfo
  {author} {\bibfnamefont {J.~E.}\ \bibnamefont {Moussa}}, \bibinfo {author}
  {\bibfnamefont {I.}~\bibnamefont {Monta\~no}}, \ and\ \bibinfo {author}
  {\bibfnamefont {R.~P.}\ \bibnamefont {Muller}},\ }\href {\doibase
  10.1103/PhysRevB.91.235318} {\bibfield  {journal} {\bibinfo  {journal} {Phys.
  Rev. B}\ }\textbf {\bibinfo {volume} {91}},\ \bibinfo {pages} {235318}
  (\bibinfo {year} {2015})}\BibitemShut {NoStop}%
\bibitem [{\citenamefont {Swift}\ \emph {et~al.}(2020)\citenamefont {Swift},
  \citenamefont {Peelaers}, \citenamefont {Mu}, \citenamefont {Morton},\ and\
  \citenamefont {Van~de Walle}}]{Swi20a}%
  \BibitemOpen
  \bibfield  {author} {\bibinfo {author} {\bibfnamefont {M.~W.}\ \bibnamefont
  {Swift}}, \bibinfo {author} {\bibfnamefont {H.}~\bibnamefont {Peelaers}},
  \bibinfo {author} {\bibfnamefont {S.}~\bibnamefont {Mu}}, \bibinfo {author}
  {\bibfnamefont {J.~J.~L.}\ \bibnamefont {Morton}}, \ and\ \bibinfo {author}
  {\bibfnamefont {C.~G.}\ \bibnamefont {Van~de Walle}},\ }\href
  {https://doi.org/10.1038/s41524-020-00448-7} {\bibfield  {journal} {\bibinfo
  {journal} {npj Computational Materials}\ }\textbf {\bibinfo {volume} {6}},\
  \bibinfo {pages} {181} (\bibinfo {year} {2020})}\BibitemShut {NoStop}%
\bibitem [{\citenamefont {Rahman}\ \emph
  {et~al.}(2009{\natexlab{b}})\citenamefont {Rahman}, \citenamefont
  {Lansbergen}, \citenamefont {Park}, \citenamefont {Verduijn}, \citenamefont
  {Klimeck}, \citenamefont {Rogge},\ and\ \citenamefont {Hollenberg}}]{Rah09a}%
  \BibitemOpen
  \bibfield  {author} {\bibinfo {author} {\bibfnamefont {R.}~\bibnamefont
  {Rahman}}, \bibinfo {author} {\bibfnamefont {G.~P.}\ \bibnamefont
  {Lansbergen}}, \bibinfo {author} {\bibfnamefont {S.~H.}\ \bibnamefont
  {Park}}, \bibinfo {author} {\bibfnamefont {J.}~\bibnamefont {Verduijn}},
  \bibinfo {author} {\bibfnamefont {G.}~\bibnamefont {Klimeck}}, \bibinfo
  {author} {\bibfnamefont {S.}~\bibnamefont {Rogge}}, \ and\ \bibinfo {author}
  {\bibfnamefont {L.~C.~L.}\ \bibnamefont {Hollenberg}},\ }\href {\doibase
  10.1103/PhysRevB.80.165314} {\bibfield  {journal} {\bibinfo  {journal} {Phys.
  Rev. B}\ }\textbf {\bibinfo {volume} {80}},\ \bibinfo {pages} {165314}
  (\bibinfo {year} {2009}{\natexlab{b}})}\BibitemShut {NoStop}%
\bibitem [{\citenamefont {Wang}\ \emph {et~al.}(2016)\citenamefont {Wang},
  \citenamefont {Tankasala}, \citenamefont {Hollenberg}, \citenamefont
  {Klimeck}, \citenamefont {Simmons},\ and\ \citenamefont {Rahman}}]{Wan16a}%
  \BibitemOpen
  \bibfield  {author} {\bibinfo {author} {\bibfnamefont {Y.}~\bibnamefont
  {Wang}}, \bibinfo {author} {\bibfnamefont {A.}~\bibnamefont {Tankasala}},
  \bibinfo {author} {\bibfnamefont {L.~C.~L.}\ \bibnamefont {Hollenberg}},
  \bibinfo {author} {\bibfnamefont {G.}~\bibnamefont {Klimeck}}, \bibinfo
  {author} {\bibfnamefont {M.~Y.}\ \bibnamefont {Simmons}}, \ and\ \bibinfo
  {author} {\bibfnamefont {R.}~\bibnamefont {Rahman}},\ }\href
  {http://dx.doi.org/10.1038/npjqi.2016.8} {\bibfield  {journal} {\bibinfo
  {journal} {Npj Quantum Information}\ }\textbf {\bibinfo {volume} {2}},\
  \bibinfo {pages} {16008} (\bibinfo {year} {2016})}\BibitemShut {NoStop}%
\bibitem [{\citenamefont {Voisin}\ \emph {et~al.}(2020)\citenamefont {Voisin},
  \citenamefont {Bocquel}, \citenamefont {Tankasala}, \citenamefont {Usman},
  \citenamefont {Salfi}, \citenamefont {Rahman}, \citenamefont {Simmons},
  \citenamefont {Hollenberg},\ and\ \citenamefont {Rogge}}]{Voi20a}%
  \BibitemOpen
  \bibfield  {author} {\bibinfo {author} {\bibfnamefont {B.}~\bibnamefont
  {Voisin}}, \bibinfo {author} {\bibfnamefont {J.}~\bibnamefont {Bocquel}},
  \bibinfo {author} {\bibfnamefont {A.}~\bibnamefont {Tankasala}}, \bibinfo
  {author} {\bibfnamefont {M.}~\bibnamefont {Usman}}, \bibinfo {author}
  {\bibfnamefont {J.}~\bibnamefont {Salfi}}, \bibinfo {author} {\bibfnamefont
  {R.}~\bibnamefont {Rahman}}, \bibinfo {author} {\bibfnamefont {M.~Y.}\
  \bibnamefont {Simmons}}, \bibinfo {author} {\bibfnamefont {L.~C.~L.}\
  \bibnamefont {Hollenberg}}, \ and\ \bibinfo {author} {\bibfnamefont
  {S.}~\bibnamefont {Rogge}},\ }\href
  {https://doi.org/10.1038/s41467-020-19835-1} {\bibfield  {journal} {\bibinfo
  {journal} {Nature Communications}\ }\textbf {\bibinfo {volume} {11}},\
  \bibinfo {pages} {6124} (\bibinfo {year} {2020})}\BibitemShut {NoStop}%
\bibitem [{\citenamefont {Verduijn}, \citenamefont {Tettamanzi},\ and\
  \citenamefont {Rogge}(2013)}]{Ver13a}%
  \BibitemOpen
  \bibfield  {author} {\bibinfo {author} {\bibfnamefont {J.}~\bibnamefont
  {Verduijn}}, \bibinfo {author} {\bibfnamefont {G.~C.}\ \bibnamefont
  {Tettamanzi}}, \ and\ \bibinfo {author} {\bibfnamefont {S.}~\bibnamefont
  {Rogge}},\ }\bibfield  {booktitle} {\emph {\bibinfo {booktitle} {Nano
  Letters}},\ }\href {\doibase https://doi.org/10.1021/nl304518v} {\bibfield
  {journal} {\bibinfo  {journal} {Nano Lett.}\ }\textbf {\bibinfo {volume}
  {13}},\ \bibinfo {pages} {1476} (\bibinfo {year} {2013})}\BibitemShut
  {NoStop}%
\bibitem [{\citenamefont {van~der Heijden}\ \emph {et~al.}(2018)\citenamefont
  {van~der Heijden}, \citenamefont {Kobayashi}, \citenamefont {House},
  \citenamefont {Salfi}, \citenamefont {Barraud}, \citenamefont
  {Lavi{\'e}ville}, \citenamefont {Simmons},\ and\ \citenamefont
  {Rogge}}]{Hei18a}%
  \BibitemOpen
  \bibfield  {author} {\bibinfo {author} {\bibfnamefont {J.}~\bibnamefont
  {van~der Heijden}}, \bibinfo {author} {\bibfnamefont {T.}~\bibnamefont
  {Kobayashi}}, \bibinfo {author} {\bibfnamefont {M.~G.}\ \bibnamefont
  {House}}, \bibinfo {author} {\bibfnamefont {J.}~\bibnamefont {Salfi}},
  \bibinfo {author} {\bibfnamefont {S.}~\bibnamefont {Barraud}}, \bibinfo
  {author} {\bibfnamefont {R.}~\bibnamefont {Lavi{\'e}ville}}, \bibinfo
  {author} {\bibfnamefont {M.~Y.}\ \bibnamefont {Simmons}}, \ and\ \bibinfo
  {author} {\bibfnamefont {S.}~\bibnamefont {Rogge}},\ }\href
  {https://advances.sciencemag.org/content/4/12/eaat9199} {\bibfield  {journal}
  {\bibinfo  {journal} {Science Advances}\ }\textbf {\bibinfo {volume} {4}}
  (\bibinfo {year} {2018})}\BibitemShut {NoStop}%
\bibitem [{\citenamefont {Verduijn}, \citenamefont {Vinet},\ and\ \citenamefont
  {Rogge}(2014)}]{Ver14a}%
  \BibitemOpen
  \bibfield  {author} {\bibinfo {author} {\bibfnamefont {J.}~\bibnamefont
  {Verduijn}}, \bibinfo {author} {\bibfnamefont {M.}~\bibnamefont {Vinet}}, \
  and\ \bibinfo {author} {\bibfnamefont {S.}~\bibnamefont {Rogge}},\ }\bibfield
   {booktitle} {\emph {\bibinfo {booktitle} {Applied Physics Letters}},\ }\href
  {\doibase https://doi.org/10.1063/1.4868423} {\bibfield  {journal} {\bibinfo
  {journal} {Applied Physics Letters}\ }\textbf {\bibinfo {volume} {104}},\
  \bibinfo {pages} {102107} (\bibinfo {year} {2014})}\BibitemShut {NoStop}%
\bibitem [{\citenamefont {Dupont-Ferrier}\ \emph {et~al.}(2013)\citenamefont
  {Dupont-Ferrier}, \citenamefont {Roche}, \citenamefont {Voisin},
  \citenamefont {Jehl}, \citenamefont {Wacquez}, \citenamefont {Vinet},
  \citenamefont {Sanquer},\ and\ \citenamefont {De~Franceschi}}]{Dup13a}%
  \BibitemOpen
  \bibfield  {author} {\bibinfo {author} {\bibfnamefont {E.}~\bibnamefont
  {Dupont-Ferrier}}, \bibinfo {author} {\bibfnamefont {B.}~\bibnamefont
  {Roche}}, \bibinfo {author} {\bibfnamefont {B.}~\bibnamefont {Voisin}},
  \bibinfo {author} {\bibfnamefont {X.}~\bibnamefont {Jehl}}, \bibinfo {author}
  {\bibfnamefont {R.}~\bibnamefont {Wacquez}}, \bibinfo {author} {\bibfnamefont
  {M.}~\bibnamefont {Vinet}}, \bibinfo {author} {\bibfnamefont
  {M.}~\bibnamefont {Sanquer}}, \ and\ \bibinfo {author} {\bibfnamefont
  {S.}~\bibnamefont {De~Franceschi}},\ }\href {\doibase
  10.1103/PhysRevLett.110.136802} {\bibfield  {journal} {\bibinfo  {journal}
  {Phys. Rev. Lett.}\ }\textbf {\bibinfo {volume} {110}},\ \bibinfo {pages}
  {136802} (\bibinfo {year} {2013})}\BibitemShut {NoStop}%
\bibitem [{\citenamefont {Fuechsle}\ \emph {et~al.}(2012)\citenamefont
  {Fuechsle}, \citenamefont {Miwa}, \citenamefont {Mahapatra}, \citenamefont
  {Ryu}, \citenamefont {Lee}, \citenamefont {Warschkow}, \citenamefont
  {Hollenberg}, \citenamefont {Klimeck},\ and\ \citenamefont
  {Simmons}}]{Fue12a}%
  \BibitemOpen
  \bibfield  {author} {\bibinfo {author} {\bibfnamefont {M.}~\bibnamefont
  {Fuechsle}}, \bibinfo {author} {\bibfnamefont {J.~A.}\ \bibnamefont {Miwa}},
  \bibinfo {author} {\bibfnamefont {S.}~\bibnamefont {Mahapatra}}, \bibinfo
  {author} {\bibfnamefont {H.}~\bibnamefont {Ryu}}, \bibinfo {author}
  {\bibfnamefont {S.}~\bibnamefont {Lee}}, \bibinfo {author} {\bibfnamefont
  {O.}~\bibnamefont {Warschkow}}, \bibinfo {author} {\bibfnamefont {L.~C.~L.}\
  \bibnamefont {Hollenberg}}, \bibinfo {author} {\bibfnamefont
  {G.}~\bibnamefont {Klimeck}}, \ and\ \bibinfo {author} {\bibfnamefont
  {M.~Y.}\ \bibnamefont {Simmons}},\ }\href
  {http://dx.doi.org/10.1038/nnano.2012.21} {\bibfield  {journal} {\bibinfo
  {journal} {Nature Nanotechnology}\ }\textbf {\bibinfo {volume} {7}},\
  \bibinfo {pages} {242} (\bibinfo {year} {2012})}\BibitemShut {NoStop}%
\bibitem [{\citenamefont {Weber}\ \emph {et~al.}(2014)\citenamefont {Weber},
  \citenamefont {Tan}, \citenamefont {Mahapatra}, \citenamefont {Watson},
  \citenamefont {Ryu}, \citenamefont {Rahman}, \citenamefont {Hollenberg},
  \citenamefont {Klimeck},\ and\ \citenamefont {Simmons}}]{Web14a}%
  \BibitemOpen
  \bibfield  {author} {\bibinfo {author} {\bibfnamefont {B.}~\bibnamefont
  {Weber}}, \bibinfo {author} {\bibfnamefont {Y.~H.~M.}\ \bibnamefont {Tan}},
  \bibinfo {author} {\bibfnamefont {S.}~\bibnamefont {Mahapatra}}, \bibinfo
  {author} {\bibfnamefont {T.~F.}\ \bibnamefont {Watson}}, \bibinfo {author}
  {\bibfnamefont {H.}~\bibnamefont {Ryu}}, \bibinfo {author} {\bibfnamefont
  {R.}~\bibnamefont {Rahman}}, \bibinfo {author} {\bibfnamefont {L.~C.~L.}\
  \bibnamefont {Hollenberg}}, \bibinfo {author} {\bibfnamefont
  {G.}~\bibnamefont {Klimeck}}, \ and\ \bibinfo {author} {\bibfnamefont
  {M.~Y.}\ \bibnamefont {Simmons}},\ }\href
  {http://dx.doi.org/10.1038/nnano.2014.63} {\bibfield  {journal} {\bibinfo
  {journal} {Nature Nanotechnology}\ }\textbf {\bibinfo {volume} {9}},\
  \bibinfo {pages} {430} (\bibinfo {year} {2014})}\BibitemShut {NoStop}%
\bibitem [{\citenamefont {Koenraad}\ and\ \citenamefont
  {Flatté}(2011)}]{Koe11a}%
  \BibitemOpen
  \bibfield  {author} {\bibinfo {author} {\bibfnamefont {P.~M.}\ \bibnamefont
  {Koenraad}}\ and\ \bibinfo {author} {\bibfnamefont {M.~E.}\ \bibnamefont
  {Flatté}},\ }\href {https://doi.org/10.1038/nmat2940} {\bibfield  {journal}
  {\bibinfo  {journal} {Nature Materials}\ }\textbf {\bibinfo {volume} {10}},\
  \bibinfo {pages} {91} (\bibinfo {year} {2011})}\BibitemShut {NoStop}%
\bibitem [{\citenamefont {Feenstra}\ \emph {et~al.}(1992)\citenamefont
  {Feenstra}, \citenamefont {Yu}, \citenamefont {Woodall}, \citenamefont
  {Kirchner}, \citenamefont {Lin},\ and\ \citenamefont {Pettit}}]{Fee92a}%
  \BibitemOpen
  \bibfield  {author} {\bibinfo {author} {\bibfnamefont {R.~M.}\ \bibnamefont
  {Feenstra}}, \bibinfo {author} {\bibfnamefont {E.~T.}\ \bibnamefont {Yu}},
  \bibinfo {author} {\bibfnamefont {J.~M.}\ \bibnamefont {Woodall}}, \bibinfo
  {author} {\bibfnamefont {P.~D.}\ \bibnamefont {Kirchner}}, \bibinfo {author}
  {\bibfnamefont {C.~L.}\ \bibnamefont {Lin}}, \ and\ \bibinfo {author}
  {\bibfnamefont {G.~D.}\ \bibnamefont {Pettit}},\ }\href {\doibase
  10.1063/1.107804} {\bibfield  {journal} {\bibinfo  {journal} {Applied Physics
  Letters}\ }\textbf {\bibinfo {volume} {61}},\ \bibinfo {pages} {795}
  (\bibinfo {year} {1992})}\BibitemShut {NoStop}%
\bibitem [{\citenamefont {Yakunin}\ \emph {et~al.}(2004)\citenamefont
  {Yakunin}, \citenamefont {Silov}, \citenamefont {Koenraad}, \citenamefont
  {Wolter}, \citenamefont {Van~Roy}, \citenamefont {De~Boeck}, \citenamefont
  {Tang},\ and\ \citenamefont {Flatt\'e}}]{Yak04a}%
  \BibitemOpen
  \bibfield  {author} {\bibinfo {author} {\bibfnamefont {A.~M.}\ \bibnamefont
  {Yakunin}}, \bibinfo {author} {\bibfnamefont {A.~Y.}\ \bibnamefont {Silov}},
  \bibinfo {author} {\bibfnamefont {P.~M.}\ \bibnamefont {Koenraad}}, \bibinfo
  {author} {\bibfnamefont {J.~H.}\ \bibnamefont {Wolter}}, \bibinfo {author}
  {\bibfnamefont {W.}~\bibnamefont {Van~Roy}}, \bibinfo {author} {\bibfnamefont
  {J.}~\bibnamefont {De~Boeck}}, \bibinfo {author} {\bibfnamefont {J.-M.}\
  \bibnamefont {Tang}}, \ and\ \bibinfo {author} {\bibfnamefont {M.~E.}\
  \bibnamefont {Flatt\'e}},\ }\href {\doibase 10.1103/PhysRevLett.92.216806}
  {\bibfield  {journal} {\bibinfo  {journal} {Phys. Rev. Lett.}\ }\textbf
  {\bibinfo {volume} {92}},\ \bibinfo {pages} {216806} (\bibinfo {year}
  {2004})}\BibitemShut {NoStop}%
\bibitem [{\citenamefont {Pitters}, \citenamefont {Piva},\ and\ \citenamefont
  {Wolkow}(2012)}]{Pit12a}%
  \BibitemOpen
  \bibfield  {author} {\bibinfo {author} {\bibfnamefont {J.~L.}\ \bibnamefont
  {Pitters}}, \bibinfo {author} {\bibfnamefont {P.~G.}\ \bibnamefont {Piva}}, \
  and\ \bibinfo {author} {\bibfnamefont {R.~A.}\ \bibnamefont {Wolkow}},\
  }\href {\doibase 10.1116/1.3694010} {\bibfield  {journal} {\bibinfo
  {journal} {Journal of Vacuum Science \& Technology B}\ }\textbf {\bibinfo
  {volume} {30}},\ \bibinfo {pages} {021806} (\bibinfo {year}
  {2012})}\BibitemShut {NoStop}%
\bibitem [{\citenamefont {Sinthiptharakoon}\ \emph {et~al.}(2013)\citenamefont
  {Sinthiptharakoon}, \citenamefont {Schofield}, \citenamefont {Studer},
  \citenamefont {Br{\'{a}}zdov{\'{a}}}, \citenamefont {Hirjibehedin},
  \citenamefont {Bowler},\ and\ \citenamefont {Curson}}]{Sin14a}%
  \BibitemOpen
  \bibfield  {author} {\bibinfo {author} {\bibfnamefont {K.}~\bibnamefont
  {Sinthiptharakoon}}, \bibinfo {author} {\bibfnamefont {S.~R.}\ \bibnamefont
  {Schofield}}, \bibinfo {author} {\bibfnamefont {P.}~\bibnamefont {Studer}},
  \bibinfo {author} {\bibfnamefont {V.}~\bibnamefont {Br{\'{a}}zdov{\'{a}}}},
  \bibinfo {author} {\bibfnamefont {C.~F.}\ \bibnamefont {Hirjibehedin}},
  \bibinfo {author} {\bibfnamefont {D.~R.}\ \bibnamefont {Bowler}}, \ and\
  \bibinfo {author} {\bibfnamefont {N.~J.}\ \bibnamefont {Curson}},\ }\href
  {\doibase 10.1088/0953-8984/26/1/012001} {\bibfield  {journal} {\bibinfo
  {journal} {Journal of Physics: Condensed Matter}\ }\textbf {\bibinfo {volume}
  {26}},\ \bibinfo {pages} {012001} (\bibinfo {year} {2013})}\BibitemShut
  {NoStop}%
\bibitem [{\citenamefont {Mol}\ \emph {et~al.}(2015)\citenamefont {Mol},
  \citenamefont {Salfi}, \citenamefont {Rahman}, \citenamefont {Hsueh},
  \citenamefont {Miwa}, \citenamefont {Klimeck}, \citenamefont {Simmons},\ and\
  \citenamefont {Rogge}}]{Mol15a}%
  \BibitemOpen
  \bibfield  {author} {\bibinfo {author} {\bibfnamefont {J.~A.}\ \bibnamefont
  {Mol}}, \bibinfo {author} {\bibfnamefont {J.}~\bibnamefont {Salfi}}, \bibinfo
  {author} {\bibfnamefont {R.}~\bibnamefont {Rahman}}, \bibinfo {author}
  {\bibfnamefont {Y.}~\bibnamefont {Hsueh}}, \bibinfo {author} {\bibfnamefont
  {J.~A.}\ \bibnamefont {Miwa}}, \bibinfo {author} {\bibfnamefont
  {G.}~\bibnamefont {Klimeck}}, \bibinfo {author} {\bibfnamefont {M.~Y.}\
  \bibnamefont {Simmons}}, \ and\ \bibinfo {author} {\bibfnamefont
  {S.}~\bibnamefont {Rogge}},\ }\href {\doibase 10.1063/1.4921640} {\bibfield
  {journal} {\bibinfo  {journal} {Applied Physics Letters}\ }\textbf {\bibinfo
  {volume} {106}},\ \bibinfo {pages} {203110} (\bibinfo {year}
  {2015})}\BibitemShut {NoStop}%
\bibitem [{\citenamefont {Salfi}\ \emph
  {et~al.}(2016{\natexlab{a}})\citenamefont {Salfi}, \citenamefont {Mol},
  \citenamefont {Rahman}, \citenamefont {Klimeck}, \citenamefont {Simmons},
  \citenamefont {Hollenberg},\ and\ \citenamefont {Rogge}}]{Sal16a}%
  \BibitemOpen
  \bibfield  {author} {\bibinfo {author} {\bibfnamefont {J.}~\bibnamefont
  {Salfi}}, \bibinfo {author} {\bibfnamefont {J.~A.}\ \bibnamefont {Mol}},
  \bibinfo {author} {\bibfnamefont {R.}~\bibnamefont {Rahman}}, \bibinfo
  {author} {\bibfnamefont {G.}~\bibnamefont {Klimeck}}, \bibinfo {author}
  {\bibfnamefont {M.~Y.}\ \bibnamefont {Simmons}}, \bibinfo {author}
  {\bibfnamefont {L.~C.~L.}\ \bibnamefont {Hollenberg}}, \ and\ \bibinfo
  {author} {\bibfnamefont {S.}~\bibnamefont {Rogge}},\ }\href
  {http://dx.doi.org/10.1038/ncomms11342} {\bibfield  {journal} {\bibinfo
  {journal} {Nature Communications}\ }\textbf {\bibinfo {volume} {7}},\
  \bibinfo {pages} {11342} (\bibinfo {year} {2016}{\natexlab{a}})}\BibitemShut
  {NoStop}%
\bibitem [{\citenamefont {Schofield}\ \emph {et~al.}(2003)\citenamefont
  {Schofield}, \citenamefont {Curson}, \citenamefont {Simmons}, \citenamefont
  {Rue\ss{}}, \citenamefont {Hallam}, \citenamefont {Oberbeck},\ and\
  \citenamefont {Clark}}]{Sch03a}%
  \BibitemOpen
  \bibfield  {author} {\bibinfo {author} {\bibfnamefont {S.~R.}\ \bibnamefont
  {Schofield}}, \bibinfo {author} {\bibfnamefont {N.~J.}\ \bibnamefont
  {Curson}}, \bibinfo {author} {\bibfnamefont {M.~Y.}\ \bibnamefont {Simmons}},
  \bibinfo {author} {\bibfnamefont {F.~J.}\ \bibnamefont {Rue\ss{}}}, \bibinfo
  {author} {\bibfnamefont {T.}~\bibnamefont {Hallam}}, \bibinfo {author}
  {\bibfnamefont {L.}~\bibnamefont {Oberbeck}}, \ and\ \bibinfo {author}
  {\bibfnamefont {R.~G.}\ \bibnamefont {Clark}},\ }\href {\doibase
  10.1103/PhysRevLett.91.136104} {\bibfield  {journal} {\bibinfo  {journal}
  {Phys. Rev. Lett.}\ }\textbf {\bibinfo {volume} {91}},\ \bibinfo {pages}
  {136104} (\bibinfo {year} {2003})}\BibitemShut {NoStop}%
\bibitem [{\citenamefont {Ng}\ \emph {et~al.}(2020)\citenamefont {Ng},
  \citenamefont {Voisin}, \citenamefont {Johnson}, \citenamefont {McCallum},
  \citenamefont {Salfi},\ and\ \citenamefont {Rogge}}]{Ng20a}%
  \BibitemOpen
  \bibfield  {author} {\bibinfo {author} {\bibfnamefont {K.~S.~H.}\
  \bibnamefont {Ng}}, \bibinfo {author} {\bibfnamefont {B.}~\bibnamefont
  {Voisin}}, \bibinfo {author} {\bibfnamefont {B.~C.}\ \bibnamefont {Johnson}},
  \bibinfo {author} {\bibfnamefont {J.~C.}\ \bibnamefont {McCallum}}, \bibinfo
  {author} {\bibfnamefont {J.}~\bibnamefont {Salfi}}, \ and\ \bibinfo {author}
  {\bibfnamefont {S.}~\bibnamefont {Rogge}},\ }\href {\doibase
  10.1021/acsnano.0c00736} {\bibfield  {journal} {\bibinfo  {journal} {ACS
  Nano}\ }\textbf {\bibinfo {volume} {14}},\ \bibinfo {pages} {9449} (\bibinfo
  {year} {2020})}\BibitemShut {NoStop}%
\bibitem [{\citenamefont {Saraiva}\ \emph {et~al.}(2016)\citenamefont
  {Saraiva}, \citenamefont {Salfi}, \citenamefont {Bocquel}, \citenamefont
  {Voisin}, \citenamefont {Rogge}, \citenamefont {Capaz}, \citenamefont
  {Calder\'on},\ and\ \citenamefont {Koiller}}]{Sar16a}%
  \BibitemOpen
  \bibfield  {author} {\bibinfo {author} {\bibfnamefont {A.~L.}\ \bibnamefont
  {Saraiva}}, \bibinfo {author} {\bibfnamefont {J.}~\bibnamefont {Salfi}},
  \bibinfo {author} {\bibfnamefont {J.}~\bibnamefont {Bocquel}}, \bibinfo
  {author} {\bibfnamefont {B.}~\bibnamefont {Voisin}}, \bibinfo {author}
  {\bibfnamefont {S.}~\bibnamefont {Rogge}}, \bibinfo {author} {\bibfnamefont
  {R.~B.}\ \bibnamefont {Capaz}}, \bibinfo {author} {\bibfnamefont {M.~J.}\
  \bibnamefont {Calder\'on}}, \ and\ \bibinfo {author} {\bibfnamefont
  {B.}~\bibnamefont {Koiller}},\ }\href {\doibase 10.1103/PhysRevB.93.045303}
  {\bibfield  {journal} {\bibinfo  {journal} {Phys. Rev. B}\ }\textbf {\bibinfo
  {volume} {93}},\ \bibinfo {pages} {045303} (\bibinfo {year}
  {2016})}\BibitemShut {NoStop}%
\bibitem [{\citenamefont {Usman}\ \emph {et~al.}(2016)\citenamefont {Usman},
  \citenamefont {Bocquel}, \citenamefont {Salfi}, \citenamefont {Voisin},
  \citenamefont {Tankasala}, \citenamefont {Rahman}, \citenamefont {Simmons},
  \citenamefont {Rogge},\ and\ \citenamefont {Hollenberg}}]{Usm16a}%
  \BibitemOpen
  \bibfield  {author} {\bibinfo {author} {\bibfnamefont {M.}~\bibnamefont
  {Usman}}, \bibinfo {author} {\bibfnamefont {J.}~\bibnamefont {Bocquel}},
  \bibinfo {author} {\bibfnamefont {J.}~\bibnamefont {Salfi}}, \bibinfo
  {author} {\bibfnamefont {B.}~\bibnamefont {Voisin}}, \bibinfo {author}
  {\bibfnamefont {A.}~\bibnamefont {Tankasala}}, \bibinfo {author}
  {\bibfnamefont {R.}~\bibnamefont {Rahman}}, \bibinfo {author} {\bibfnamefont
  {M.~Y.}\ \bibnamefont {Simmons}}, \bibinfo {author} {\bibfnamefont
  {S.}~\bibnamefont {Rogge}}, \ and\ \bibinfo {author} {\bibfnamefont
  {L.~C.~L.}\ \bibnamefont {Hollenberg}},\ }\href {\doibase
  10.1038/nnano.2016.83} {\bibfield  {journal} {\bibinfo  {journal} {Nature
  Nanotechnology}\ }\textbf {\bibinfo {volume} {11}},\ \bibinfo {pages} {1}
  (\bibinfo {year} {2016})}\BibitemShut {NoStop}%
\bibitem [{\citenamefont {Br\'azdov\'a}\ \emph {et~al.}(2017)\citenamefont
  {Br\'azdov\'a}, \citenamefont {Bowler}, \citenamefont {Sinthiptharakoon},
  \citenamefont {Studer}, \citenamefont {Rahnejat}, \citenamefont {Curson},
  \citenamefont {Schofield},\ and\ \citenamefont {Fisher}}]{Bra17a}%
  \BibitemOpen
  \bibfield  {author} {\bibinfo {author} {\bibfnamefont {V.}~\bibnamefont
  {Br\'azdov\'a}}, \bibinfo {author} {\bibfnamefont {D.~R.}\ \bibnamefont
  {Bowler}}, \bibinfo {author} {\bibfnamefont {K.}~\bibnamefont
  {Sinthiptharakoon}}, \bibinfo {author} {\bibfnamefont {P.}~\bibnamefont
  {Studer}}, \bibinfo {author} {\bibfnamefont {A.}~\bibnamefont {Rahnejat}},
  \bibinfo {author} {\bibfnamefont {N.~J.}\ \bibnamefont {Curson}}, \bibinfo
  {author} {\bibfnamefont {S.~R.}\ \bibnamefont {Schofield}}, \ and\ \bibinfo
  {author} {\bibfnamefont {A.~J.}\ \bibnamefont {Fisher}},\ }\href {\doibase
  10.1103/PhysRevB.95.075408} {\bibfield  {journal} {\bibinfo  {journal} {Phys.
  Rev. B}\ }\textbf {\bibinfo {volume} {95}},\ \bibinfo {pages} {075408}
  (\bibinfo {year} {2017})}\BibitemShut {NoStop}%
\bibitem [{\citenamefont {Georgescu}, \citenamefont {Ashhab},\ and\
  \citenamefont {Nori}(2014)}]{Geo14a}%
  \BibitemOpen
  \bibfield  {author} {\bibinfo {author} {\bibfnamefont {I.~M.}\ \bibnamefont
  {Georgescu}}, \bibinfo {author} {\bibfnamefont {S.}~\bibnamefont {Ashhab}}, \
  and\ \bibinfo {author} {\bibfnamefont {F.}~\bibnamefont {Nori}},\ }\href
  {\doibase 10.1103/RevModPhys.86.153} {\bibfield  {journal} {\bibinfo
  {journal} {Rev. Mod. Phys.}\ }\textbf {\bibinfo {volume} {86}},\ \bibinfo
  {pages} {153} (\bibinfo {year} {2014})}\BibitemShut {NoStop}%
\bibitem [{\citenamefont {Wellard}\ and\ \citenamefont
  {Hollenberg}(2005)}]{Wel05a}%
  \BibitemOpen
  \bibfield  {author} {\bibinfo {author} {\bibfnamefont {C.~J.}\ \bibnamefont
  {Wellard}}\ and\ \bibinfo {author} {\bibfnamefont {L.~C.~L.}\ \bibnamefont
  {Hollenberg}},\ }\href {\doibase 10.1103/PhysRevB.72.085202} {\bibfield
  {journal} {\bibinfo  {journal} {Phys. Rev. B}\ }\textbf {\bibinfo {volume}
  {72}},\ \bibinfo {pages} {085202} (\bibinfo {year} {2005})}\BibitemShut
  {NoStop}%
\bibitem [{\citenamefont {Madzik}\ \emph {et~al.}(2021)\citenamefont {Madzik},
  \citenamefont {Laucht}, \citenamefont {Hudson}, \citenamefont {Jakob},
  \citenamefont {Johnson}, \citenamefont {Jamieson}, \citenamefont {Itoh},
  \citenamefont {Dzurak},\ and\ \citenamefont {Morello}}]{Mad20a}%
  \BibitemOpen
  \bibfield  {author} {\bibinfo {author} {\bibfnamefont {M.~T.}\ \bibnamefont
  {Madzik}}, \bibinfo {author} {\bibfnamefont {A.}~\bibnamefont {Laucht}},
  \bibinfo {author} {\bibfnamefont {F.~E.}\ \bibnamefont {Hudson}}, \bibinfo
  {author} {\bibfnamefont {A.~M.}\ \bibnamefont {Jakob}}, \bibinfo {author}
  {\bibfnamefont {B.~C.}\ \bibnamefont {Johnson}}, \bibinfo {author}
  {\bibfnamefont {D.~N.}\ \bibnamefont {Jamieson}}, \bibinfo {author}
  {\bibfnamefont {K.~M.}\ \bibnamefont {Itoh}}, \bibinfo {author}
  {\bibfnamefont {A.~S.}\ \bibnamefont {Dzurak}}, \ and\ \bibinfo {author}
  {\bibfnamefont {A.}~\bibnamefont {Morello}},\ }\href
  {https://doi.org/10.1038/s41467-020-20424-5} {\bibfield  {journal} {\bibinfo
  {journal} {Nature Communications}\ }\textbf {\bibinfo {volume} {12}},\
  \bibinfo {pages} {181} (\bibinfo {year} {2021})}\BibitemShut {NoStop}%
\bibitem [{\citenamefont {Kobayashi}\ \emph {et~al.}(2021)\citenamefont
  {Kobayashi}, \citenamefont {Salfi}, \citenamefont {Chua}, \citenamefont
  {van~der Heijden}, \citenamefont {House}, \citenamefont {Culcer},
  \citenamefont {Hutchison}, \citenamefont {Johnson}, \citenamefont {McCallum},
  \citenamefont {Riemann}, \citenamefont {Abrosimov}, \citenamefont {Becker},
  \citenamefont {Pohl}, \citenamefont {Simmons},\ and\ \citenamefont
  {Rogge}}]{Kob20a}%
  \BibitemOpen
  \bibfield  {author} {\bibinfo {author} {\bibfnamefont {T.}~\bibnamefont
  {Kobayashi}}, \bibinfo {author} {\bibfnamefont {J.}~\bibnamefont {Salfi}},
  \bibinfo {author} {\bibfnamefont {C.}~\bibnamefont {Chua}}, \bibinfo {author}
  {\bibfnamefont {J.}~\bibnamefont {van~der Heijden}}, \bibinfo {author}
  {\bibfnamefont {M.~G.}\ \bibnamefont {House}}, \bibinfo {author}
  {\bibfnamefont {D.}~\bibnamefont {Culcer}}, \bibinfo {author} {\bibfnamefont
  {W.~D.}\ \bibnamefont {Hutchison}}, \bibinfo {author} {\bibfnamefont {B.~C.}\
  \bibnamefont {Johnson}}, \bibinfo {author} {\bibfnamefont {J.~C.}\
  \bibnamefont {McCallum}}, \bibinfo {author} {\bibfnamefont {H.}~\bibnamefont
  {Riemann}}, \bibinfo {author} {\bibfnamefont {N.~V.}\ \bibnamefont
  {Abrosimov}}, \bibinfo {author} {\bibfnamefont {P.}~\bibnamefont {Becker}},
  \bibinfo {author} {\bibfnamefont {H.-J.}\ \bibnamefont {Pohl}}, \bibinfo
  {author} {\bibfnamefont {M.~Y.}\ \bibnamefont {Simmons}}, \ and\ \bibinfo
  {author} {\bibfnamefont {S.}~\bibnamefont {Rogge}},\ }\href {\doibase
  https://doi.org/10.1038/s41563-020-0743-3} {\bibfield  {journal} {\bibinfo
  {journal} {Nature Materials}\ }\textbf {\bibinfo {volume} {20}},\ \bibinfo
  {pages} {38} (\bibinfo {year} {2021})}\BibitemShut {NoStop}%
\bibitem [{\citenamefont {Asaad}\ \emph {et~al.}(2020)\citenamefont {Asaad},
  \citenamefont {Mourik}, \citenamefont {Joecker}, \citenamefont {Johnson},
  \citenamefont {Baczewski}, \citenamefont {Firgau}, \citenamefont {Madzik},
  \citenamefont {Schmitt}, \citenamefont {Pla}, \citenamefont {Hudson},
  \citenamefont {Itoh}, \citenamefont {McCallum}, \citenamefont {Dzurak},
  \citenamefont {Laucht},\ and\ \citenamefont {Morello}}]{Asa20a}%
  \BibitemOpen
  \bibfield  {author} {\bibinfo {author} {\bibfnamefont {S.}~\bibnamefont
  {Asaad}}, \bibinfo {author} {\bibfnamefont {V.}~\bibnamefont {Mourik}},
  \bibinfo {author} {\bibfnamefont {B.}~\bibnamefont {Joecker}}, \bibinfo
  {author} {\bibfnamefont {M.~A.~I.}\ \bibnamefont {Johnson}}, \bibinfo
  {author} {\bibfnamefont {A.~D.}\ \bibnamefont {Baczewski}}, \bibinfo {author}
  {\bibfnamefont {H.~R.}\ \bibnamefont {Firgau}}, \bibinfo {author}
  {\bibfnamefont {M.~T.}\ \bibnamefont {Madzik}}, \bibinfo {author}
  {\bibfnamefont {V.}~\bibnamefont {Schmitt}}, \bibinfo {author} {\bibfnamefont
  {J.~J.}\ \bibnamefont {Pla}}, \bibinfo {author} {\bibfnamefont {F.~E.}\
  \bibnamefont {Hudson}}, \bibinfo {author} {\bibfnamefont {K.~M.}\
  \bibnamefont {Itoh}}, \bibinfo {author} {\bibfnamefont {J.~C.}\ \bibnamefont
  {McCallum}}, \bibinfo {author} {\bibfnamefont {A.~S.}\ \bibnamefont
  {Dzurak}}, \bibinfo {author} {\bibfnamefont {A.}~\bibnamefont {Laucht}}, \
  and\ \bibinfo {author} {\bibfnamefont {A.}~\bibnamefont {Morello}},\ }\href
  {\doibase https://doi.org/10.1038/s41586-020-2057-7} {\bibfield  {journal}
  {\bibinfo  {journal} {Nature}\ }\textbf {\bibinfo {volume} {579}},\ \bibinfo
  {pages} {205} (\bibinfo {year} {2020})}\BibitemShut {NoStop}%
\bibitem [{\citenamefont {Yin}\ \emph {et~al.}(2013)\citenamefont {Yin},
  \citenamefont {Rancic}, \citenamefont {de~Boo}, \citenamefont {Stavrias},
  \citenamefont {McCallum}, \citenamefont {Sellars},\ and\ \citenamefont
  {Rogge}}]{Yin13a}%
  \BibitemOpen
  \bibfield  {author} {\bibinfo {author} {\bibfnamefont {C.}~\bibnamefont
  {Yin}}, \bibinfo {author} {\bibfnamefont {M.}~\bibnamefont {Rancic}},
  \bibinfo {author} {\bibfnamefont {G.~G.}\ \bibnamefont {de~Boo}}, \bibinfo
  {author} {\bibfnamefont {N.}~\bibnamefont {Stavrias}}, \bibinfo {author}
  {\bibfnamefont {J.~C.}\ \bibnamefont {McCallum}}, \bibinfo {author}
  {\bibfnamefont {M.~J.}\ \bibnamefont {Sellars}}, \ and\ \bibinfo {author}
  {\bibfnamefont {S.}~\bibnamefont {Rogge}},\ }\href {\doibase
  https://doi.org/10.1038/nature12081} {\bibfield  {journal} {\bibinfo
  {journal} {Nature}\ }\textbf {\bibinfo {volume} {497}},\ \bibinfo {pages}
  {91} (\bibinfo {year} {2013})}\BibitemShut {NoStop}%
\bibitem [{\citenamefont {Bergeron}\ \emph {et~al.}(2020)\citenamefont
  {Bergeron}, \citenamefont {Chartrand}, \citenamefont {Kurkjian},
  \citenamefont {Morse}, \citenamefont {Riemann}, \citenamefont {Abrosimov},
  \citenamefont {Becker}, \citenamefont {Pohl}, \citenamefont {Thewalt},\ and\
  \citenamefont {Simmons}}]{Ber20a}%
  \BibitemOpen
  \bibfield  {author} {\bibinfo {author} {\bibfnamefont {L.}~\bibnamefont
  {Bergeron}}, \bibinfo {author} {\bibfnamefont {C.}~\bibnamefont {Chartrand}},
  \bibinfo {author} {\bibfnamefont {A.~T.~K.}\ \bibnamefont {Kurkjian}},
  \bibinfo {author} {\bibfnamefont {K.~J.}\ \bibnamefont {Morse}}, \bibinfo
  {author} {\bibfnamefont {H.}~\bibnamefont {Riemann}}, \bibinfo {author}
  {\bibfnamefont {N.~V.}\ \bibnamefont {Abrosimov}}, \bibinfo {author}
  {\bibfnamefont {P.}~\bibnamefont {Becker}}, \bibinfo {author} {\bibfnamefont
  {H.~J.}\ \bibnamefont {Pohl}}, \bibinfo {author} {\bibfnamefont {M.~L.~W.}\
  \bibnamefont {Thewalt}}, \ and\ \bibinfo {author} {\bibfnamefont
  {S.}~\bibnamefont {Simmons}},\ }\href {\doibase 10.1103/PRXQuantum.1.020301}
  {\bibfield  {journal} {\bibinfo  {journal} {PRX Quantum}\ }\textbf {\bibinfo
  {volume} {1}},\ \bibinfo {pages} {020301} (\bibinfo {year}
  {2020})}\BibitemShut {NoStop}%
\bibitem [{\citenamefont {Bernevig}\ and\ \citenamefont
  {Zhang}(2005)}]{Ber05a}%
  \BibitemOpen
  \bibfield  {author} {\bibinfo {author} {\bibfnamefont {B.~A.}\ \bibnamefont
  {Bernevig}}\ and\ \bibinfo {author} {\bibfnamefont {S.-C.}\ \bibnamefont
  {Zhang}},\ }\href {\doibase 10.1103/PhysRevB.71.035303} {\bibfield  {journal}
  {\bibinfo  {journal} {Phys. Rev. B}\ }\textbf {\bibinfo {volume} {71}},\
  \bibinfo {pages} {035303} (\bibinfo {year} {2005})}\BibitemShut {NoStop}%
\bibitem [{\citenamefont {Petersson}\ \emph {et~al.}(2012)\citenamefont
  {Petersson}, \citenamefont {McFaul}, \citenamefont {Schroer}, \citenamefont
  {Jung}, \citenamefont {Taylor}, \citenamefont {Houck},\ and\ \citenamefont
  {Petta}}]{Pet12a}%
  \BibitemOpen
  \bibfield  {author} {\bibinfo {author} {\bibfnamefont {K.~D.}\ \bibnamefont
  {Petersson}}, \bibinfo {author} {\bibfnamefont {L.~W.}\ \bibnamefont
  {McFaul}}, \bibinfo {author} {\bibfnamefont {M.~D.}\ \bibnamefont {Schroer}},
  \bibinfo {author} {\bibfnamefont {M.}~\bibnamefont {Jung}}, \bibinfo {author}
  {\bibfnamefont {J.~M.}\ \bibnamefont {Taylor}}, \bibinfo {author}
  {\bibfnamefont {A.~A.}\ \bibnamefont {Houck}}, \ and\ \bibinfo {author}
  {\bibfnamefont {J.~R.}\ \bibnamefont {Petta}},\ }\href
  {https://doi.org/10.1038/nature11559} {\bibfield  {journal} {\bibinfo
  {journal} {Nature}\ }\textbf {\bibinfo {volume} {490}},\ \bibinfo {pages}
  {380} (\bibinfo {year} {2012})}\BibitemShut {NoStop}%
\bibitem [{\citenamefont {Salfi}\ \emph
  {et~al.}(2016{\natexlab{b}})\citenamefont {Salfi}, \citenamefont {Mol},
  \citenamefont {Culcer},\ and\ \citenamefont {Rogge}}]{Sal16b}%
  \BibitemOpen
  \bibfield  {author} {\bibinfo {author} {\bibfnamefont {J.}~\bibnamefont
  {Salfi}}, \bibinfo {author} {\bibfnamefont {J.~A.}\ \bibnamefont {Mol}},
  \bibinfo {author} {\bibfnamefont {D.}~\bibnamefont {Culcer}}, \ and\ \bibinfo
  {author} {\bibfnamefont {S.}~\bibnamefont {Rogge}},\ }\href {\doibase
  10.1103/PhysRevLett.116.246801} {\bibfield  {journal} {\bibinfo  {journal}
  {Phys. Rev. Lett.}\ }\textbf {\bibinfo {volume} {116}},\ \bibinfo {pages}
  {246801} (\bibinfo {year} {2016}{\natexlab{b}})}\BibitemShut {NoStop}%
\bibitem [{\citenamefont {Mourik}\ \emph {et~al.}(2018)\citenamefont {Mourik},
  \citenamefont {Asaad}, \citenamefont {Firgau}, \citenamefont {Pla},
  \citenamefont {Holmes}, \citenamefont {Milburn}, \citenamefont {McCallum},\
  and\ \citenamefont {Morello}}]{Mou18a}%
  \BibitemOpen
  \bibfield  {author} {\bibinfo {author} {\bibfnamefont {V.}~\bibnamefont
  {Mourik}}, \bibinfo {author} {\bibfnamefont {S.}~\bibnamefont {Asaad}},
  \bibinfo {author} {\bibfnamefont {H.}~\bibnamefont {Firgau}}, \bibinfo
  {author} {\bibfnamefont {J.~J.}\ \bibnamefont {Pla}}, \bibinfo {author}
  {\bibfnamefont {C.}~\bibnamefont {Holmes}}, \bibinfo {author} {\bibfnamefont
  {G.~J.}\ \bibnamefont {Milburn}}, \bibinfo {author} {\bibfnamefont {J.~C.}\
  \bibnamefont {McCallum}}, \ and\ \bibinfo {author} {\bibfnamefont
  {A.}~\bibnamefont {Morello}},\ }\href {\doibase 10.1103/PhysRevE.98.042206}
  {\bibfield  {journal} {\bibinfo  {journal} {Phys. Rev. E}\ }\textbf {\bibinfo
  {volume} {98}},\ \bibinfo {pages} {042206} (\bibinfo {year}
  {2018})}\BibitemShut {NoStop}%
\bibitem [{\citenamefont {Sieberer}\ \emph {et~al.}(2019)\citenamefont
  {Sieberer}, \citenamefont {Olsacher}, \citenamefont {Elben}, \citenamefont
  {Heyl}, \citenamefont {Hauke}, \citenamefont {Haake},\ and\ \citenamefont
  {Zoller}}]{Sie19a}%
  \BibitemOpen
  \bibfield  {author} {\bibinfo {author} {\bibfnamefont {L.~M.}\ \bibnamefont
  {Sieberer}}, \bibinfo {author} {\bibfnamefont {T.}~\bibnamefont {Olsacher}},
  \bibinfo {author} {\bibfnamefont {A.}~\bibnamefont {Elben}}, \bibinfo
  {author} {\bibfnamefont {M.}~\bibnamefont {Heyl}}, \bibinfo {author}
  {\bibfnamefont {P.}~\bibnamefont {Hauke}}, \bibinfo {author} {\bibfnamefont
  {F.}~\bibnamefont {Haake}}, \ and\ \bibinfo {author} {\bibfnamefont
  {P.}~\bibnamefont {Zoller}},\ }\href {\doibase
  https://doi.org/10.1038/s41534-019-0192-5} {\bibfield  {journal} {\bibinfo
  {journal} {npj Quantum Information}\ }\textbf {\bibinfo {volume} {5}},\
  \bibinfo {pages} {1} (\bibinfo {year} {2019})}\BibitemShut {NoStop}%
\bibitem [{\citenamefont {Kurizki}\ \emph {et~al.}(2015)\citenamefont
  {Kurizki}, \citenamefont {Bertet}, \citenamefont {Kubo}, \citenamefont
  {M{\o}lmer}, \citenamefont {Petrosyan}, \citenamefont {Rabl},\ and\
  \citenamefont {Schmiedmayer}}]{Kur15a}%
  \BibitemOpen
  \bibfield  {author} {\bibinfo {author} {\bibfnamefont {G.}~\bibnamefont
  {Kurizki}}, \bibinfo {author} {\bibfnamefont {P.}~\bibnamefont {Bertet}},
  \bibinfo {author} {\bibfnamefont {Y.}~\bibnamefont {Kubo}}, \bibinfo {author}
  {\bibfnamefont {K.}~\bibnamefont {M{\o}lmer}}, \bibinfo {author}
  {\bibfnamefont {D.}~\bibnamefont {Petrosyan}}, \bibinfo {author}
  {\bibfnamefont {P.}~\bibnamefont {Rabl}}, \ and\ \bibinfo {author}
  {\bibfnamefont {J.}~\bibnamefont {Schmiedmayer}},\ }\href {\doibase
  10.1073/pnas.1419326112} {\bibfield  {journal} {\bibinfo  {journal}
  {Proceedings of the National Academy of Sciences}\ }\textbf {\bibinfo
  {volume} {112}},\ \bibinfo {pages} {3866} (\bibinfo {year}
  {2015})}\BibitemShut {NoStop}%
\bibitem [{\citenamefont {Ruskov}\ and\ \citenamefont {Tahan}(2013)}]{Rus13a}%
  \BibitemOpen
  \bibfield  {author} {\bibinfo {author} {\bibfnamefont {R.}~\bibnamefont
  {Ruskov}}\ and\ \bibinfo {author} {\bibfnamefont {C.}~\bibnamefont {Tahan}},\
  }\href {\doibase 10.1103/PhysRevB.88.064308} {\bibfield  {journal} {\bibinfo
  {journal} {Phys. Rev. B}\ }\textbf {\bibinfo {volume} {88}},\ \bibinfo
  {pages} {064308} (\bibinfo {year} {2013})}\BibitemShut {NoStop}%
\bibitem [{\citenamefont {Mol}\ \emph {et~al.}(2013)\citenamefont {Mol},
  \citenamefont {Salfi}, \citenamefont {Miwa}, \citenamefont {Simmons},\ and\
  \citenamefont {Rogge}}]{Mol13a}%
  \BibitemOpen
  \bibfield  {author} {\bibinfo {author} {\bibfnamefont {J.~A.}\ \bibnamefont
  {Mol}}, \bibinfo {author} {\bibfnamefont {J.}~\bibnamefont {Salfi}}, \bibinfo
  {author} {\bibfnamefont {J.~A.}\ \bibnamefont {Miwa}}, \bibinfo {author}
  {\bibfnamefont {M.~Y.}\ \bibnamefont {Simmons}}, \ and\ \bibinfo {author}
  {\bibfnamefont {S.}~\bibnamefont {Rogge}},\ }\href {\doibase
  10.1103/PhysRevB.87.245417} {\bibfield  {journal} {\bibinfo  {journal} {Phys.
  Rev. B}\ }\textbf {\bibinfo {volume} {87}},\ \bibinfo {pages} {245417}
  (\bibinfo {year} {2013})}\BibitemShut {NoStop}%
\bibitem [{\citenamefont {Zhong}\ \emph {et~al.}(2015)\citenamefont {Zhong},
  \citenamefont {Hedges}, \citenamefont {Ahlefeldt}, \citenamefont
  {Bartholomew}, \citenamefont {Beavan}, \citenamefont {Wittig}, \citenamefont
  {Longdell},\ and\ \citenamefont {Sellars}}]{Zho15a}%
  \BibitemOpen
  \bibfield  {author} {\bibinfo {author} {\bibfnamefont {M.}~\bibnamefont
  {Zhong}}, \bibinfo {author} {\bibfnamefont {M.~P.}\ \bibnamefont {Hedges}},
  \bibinfo {author} {\bibfnamefont {R.~L.}\ \bibnamefont {Ahlefeldt}}, \bibinfo
  {author} {\bibfnamefont {J.~G.}\ \bibnamefont {Bartholomew}}, \bibinfo
  {author} {\bibfnamefont {S.~E.}\ \bibnamefont {Beavan}}, \bibinfo {author}
  {\bibfnamefont {S.~M.}\ \bibnamefont {Wittig}}, \bibinfo {author}
  {\bibfnamefont {J.~J.}\ \bibnamefont {Longdell}}, \ and\ \bibinfo {author}
  {\bibfnamefont {M.~J.}\ \bibnamefont {Sellars}},\ }\href {\doibase
  https://doi.org/10.1038/nature14025} {\bibfield  {journal} {\bibinfo
  {journal} {Nature}\ }\textbf {\bibinfo {volume} {517}},\ \bibinfo {pages}
  {177} (\bibinfo {year} {2015})}\BibitemShut {NoStop}%
\bibitem [{\citenamefont {Weiss}\ \emph {et~al.}(2021)\citenamefont {Weiss},
  \citenamefont {Gritsch}, \citenamefont {Merkel},\ and\ \citenamefont
  {Reiserer}}]{Wei20pp}%
  \BibitemOpen
  \bibfield  {author} {\bibinfo {author} {\bibfnamefont {L.}~\bibnamefont
  {Weiss}}, \bibinfo {author} {\bibfnamefont {A.}~\bibnamefont {Gritsch}},
  \bibinfo {author} {\bibfnamefont {B.}~\bibnamefont {Merkel}}, \ and\ \bibinfo
  {author} {\bibfnamefont {A.}~\bibnamefont {Reiserer}},\ }\href {\doibase
  10.1364/OPTICA.413330} {\bibfield  {journal} {\bibinfo  {journal} {Optica}\
  }\textbf {\bibinfo {volume} {8}},\ \bibinfo {pages} {40} (\bibinfo {year}
  {2021})}\BibitemShut {NoStop}%
\end{thebibliography}%

\end{document}